\documentclass{aims} 
\usepackage{amsmath}
\usepackage{paralist}
\usepackage[misc]{ifsym}
\usepackage{epsfig} 
\usepackage{epstopdf} 
\usepackage[colorlinks=true]{hyperref}
\usepackage{microtype}
\usepackage{graphicx}
\usepackage{subfigure}
\usepackage{booktabs} 
\usepackage{multirow}
\usepackage{amssymb}
\usepackage{mathtools}
\usepackage{amsthm}
\usepackage{algorithm}
\usepackage{algorithmic}
\hypersetup{urlcolor=blue, citecolor=red}
\allowdisplaybreaks

\def\currentvolume{X}
 \def\currentissue{X}
  \def\currentyear{200X}
   \def\currentmonth{XX}
    \def\ppages{X-XX}
     \def\DOI{10.3934/xx.xxxxxxx}

\theoremstyle{definition}

\title[Pseudo-Hamiltonian System Identification]
{Pseudo-Hamiltonian System Identification} 

\author[Sigurd Holmsen, Sølve Eidnes and Signe Riemer-Sørensen]{}

\subjclass{Primary: 34A55, 37M10; Secondary: 37J99.}
\keywords{Physics-informed machine learning, hybrid modelling, dynamical systems, system identification.}

\thanks{This research was supported by the Research Council of Norway and the industry partners Elkem, Eramet Norway, Equinor, BP, Subsea7, Kongsberg Maritime, Aker Solutions and Veas, through the projects BigDataMine (project no. 309691), PRAI (Prediction of Riser-response by Artificial Intelligence) (project no. 308832) and PhysML: Structure-based machine learning for physical systems (project no. 338779).}

\thanks{$^*$Corresponding author}

\begin{document}
\maketitle

\centerline{\scshape
Sigurd Holmsen$^{{\href{mailto:sigurd.holmsen@norges-bank.no}{\textrm{\Letter}}}1}$,
Sølve Eidnes$^{{\href{mailto:solve.eidnes@sintef.no}{\textrm{\Letter}}}*2}$
and Signe Riemer-Sørensen$^{{\href{mailto:signe.riemer-sorensen@sintef.no}{\textrm{\Letter}}}2}$}

\medskip

{\footnotesize
 \centerline{$^1$Norges Bank, 0107 Oslo, Norway}
} 

\medskip

{\footnotesize
 \centerline{$^2$Department of Mathematics and Cybernetics, SINTEF Digital, 0373 Oslo, Norway}
}

\bigskip



\begin{abstract}
Identifying the underlying dynamics of physical systems can be challenging when only provided with observational data. In this work, we consider systems that can be modelled as first-order ordinary differential equations. By assuming a certain pseudo-Hamiltonian formulation, we are able to learn the analytic terms of internal dynamics even if the model is trained on data where the system is affected by unknown damping and external disturbances. In cases where it is difficult to find analytic terms for the disturbances, a hybrid model that uses a neural network to learn these can still accurately identify the dynamics of the system as if under ideal conditions. This makes the models applicable in some situations where other system identification models fail. Furthermore, we propose to use a fourth-order symmetric integration scheme in the loss function and avoid actual integration in the training, and demonstrate on varied examples how this leads to increased performance on noisy data.
\end{abstract}


\section{Introduction}
Ever since the concept of describing physical systems by differential equations was introduced with the invention of calculus, humans have sought to identify the differential equations that most accurately describe a given system. The identification of the correct mathematical terms to include has historically been based on a combination of qualitative analysis, i.e.\ assumptions about the forces and physical laws that affect the system, and quantitative analysis, i.e.\ evaluation of data collected from experiments. As the capabilities of computers to store and handle large quantities of data increase, together with decreasing costs of sensors and computers, quantitative analysis can take an increasingly dominant role in system identification.

Two clear and important advantages of identifying analytic terms for a system, rather than learning a black-box model, stand out: For one, analytic terms can give insight into the process under consideration that may increase understanding and spur further developments. Secondly, accurate analytic models are less likely to suffer from the poor extrapolation abilities that make most machine learning models not applicable outside the range of the training data.

Recently, there has been a number of works on neural network models for dynamical systems. These aim at training a model to approximate the right-hand side of a first-order ordinary differential equation (ODE) system
\begin{equation}\label{eq:ode}
    \dot{x} = g(x,t), \quad x \in \mathbb{R}^d, \quad t \in \mathbb{R}.
\end{equation}
Physics-informed machine learning is a field motivated by the idea that prior assumptions and knowledge should be imposed on the machine learning model so that it does not have to be relearned \cite{Raissi2019physics, Karniadakis2021physics, Willard2022integrating}. This is especially relevant for learning dynamical systems. Much of the literature has focused on Hamiltonian or Lagrangian formulations of \eqref{eq:ode}, beginning with \cite{Greydanus2019hamiltonian, Chen2020Symplectic, Cranmer2020lagrangian}. In \cite{Duong2021hamiltonian, Eidnes2022port}, it is demonstrated how a \textit{pseudo-Hamiltonian} formulation can be utilized in the model structure to separate the internal dynamics and external forces acting on the system. This pseudo-Hamiltonian formulation is a generalization of the port-Hamiltonian formulation \cite{van2014port}, which is again a generalization of the Hamiltonian formulation. The key innovation of these pseudo-Hamiltonian neural networks (PHNN) is their ability to learn a model for the full dynamical system \eqref{eq:ode} from data while simultaneously learning models for the external forces acting on the system. 
The external forces can be removed from the model after training, and thus a model trained under suboptimal conditions learns the properties of the system itself and hence can be used to model the system as if under optimal conditions.

This addresses a major limitation of system identification. Most frameworks struggle to learn a system when the data is sampled under suboptimal conditions, e.g.\ when there is some external disturbance affecting the dynamics; the frameworks cannot disentangle the disturbances from the system itself. The motivating idea behind this paper is to use the pseudo-Hamiltonian formulation to separate out external forces while identifying analytic terms for the internal dynamics. This can make system identification applicable for more complicated and realistic systems than those frequently studied in the literature.

Moreover, we tackle the problem of how to learn models of systems described by differential equations when we cannot assume to have data on derivatives. This has been extensively studied for neural network models \cite{Jin2020sympnets, Matsubara2020deep, Zhu2022numerical}, but less so for system identification models. We propose here to train on a numerical integration scheme without actually integrating so that we can even out the noise by including two data points in the discretization of \eqref{eq:ode}.

This paper aims to advance the field of system identification toward a wider applicability to real-world problems through two main contributions:
\begin{itemize}
    \item By assuming a pseudo-Hamiltonian formulation of the system, we propose models that can be used to learn conservation laws, damping and external forces simultaneously, where external forces can be learned by a neural network model if they are difficult to express analytically.
    \item We argue for using symmetric numerical integrators in the training of the model and demonstrate superior performance over existing methods, especially on noisy data.
\end{itemize}
The implementation of the PHSI models is done in Python and builds on the \texttt{phlearn} package introduced in \cite{Eidnes2022port}. Code to reproduce numerical experiments from the paper is published at \url{https://github.com/SINTEF/PHSI}.

\section{Related work}\label{sec:related}

This paper stands on two broad shoulders: The first is the long history of and extensive literature on system identification, see e.g.\ \cite{Ljung1987system}, and especially recent developments that involve applying machine learning techniques and sparse representations, as made popular by the sparse identification of nonlinear dynamics (SINDy) framework \cite{brunton2016sindy}. The second consists of recent developments in data-driven models of dynamical systems informed of a certain structure and studies on numerical integrators in the training of these models, where most of the literature to date has been on neural network models. 

The SINDy approach for system identification has been extensively studied and extended to a wide area of applications \cite{Rackauckas2020universal, Klus2020data}. These include areas where (pseudo-)Hamiltonian formulations are of interest, like control theory \cite{Kaiser2018sparse, Fasel2022ensemble} and partial differential equations \cite{Rudy2017data, Schaeffer2017learning, Kaheman2020sindy}, and hence our methodology could also be applicable here. The SINDy approach leverages $L_1$-regularized regression to prune the space of terms to include in the analytic expression for \eqref{eq:ode}. This makes it less dependent on large amounts of data and more applicable for nonlinear systems than genetic algorithms \cite{Koza1994genetic} and symbolic regression \cite{Schmidt2009distilling}. However, there has been recent advances exploiting symbolic regression in different ways such as using filters for the derivatives and combining with a symbolic neural network \cite{Long2019pde}, learning a graph neural network before applying symbolic regression \cite{Cranmer2020discovering}, and AI Feynman \cite{Udrescu2020ai}, which provides a framework for system identification in a step-wise and recursive procedure that involves detecting structures in the system to reduce the search space. Although terms involving derivatives are not considered in \cite{Udrescu2020ai}, this could be incorporated to allow for discovering differential equations. Hence our proposed method could be integrated as a step in a framework similar to AI Feynman, where the identified pseudo-Hamiltonian structure would indicate which terms to include in the search space. In that case, it could be the next step after detecting conservation laws and neural network models of conserved quantities, e.g.\ as done in \cite{Liu2021machine, Alet2021noether, Matsubara2022finde}.

Hamiltonian neural networks (HNN) have received considerable attention since their introduction in \cite{Greydanus2019hamiltonian}, resulting in several extensions and generalizations \cite{Chen2020Symplectic, Finzi2020simplifying, Jin2022learning}. Especially relevant for this paper are the generalizations of the Hamiltonian formulation allowing for damping, disturbances acting on the system, or interaction with other systems. In addition to those that use the term port-Hamiltonian \cite{Desai2021port, Duong2021hamiltonian} or pseudo-Hamiltonian \cite{Eidnes2022port, Eidnes2023pseudo}, several papers consider a specific generalization of Hamiltonian systems that fall under the definition of pseudo-Hamiltonian systems applied in this paper, whether they aim to learn external forces \cite{Duong2022adaptive} or not \cite{Zhong2020symplectic, Zhong2020dissipative, Zhang2022gfinns}. Our motivation for the pseudo-Hamiltonian formulation comes from the long history of modeling mechanical systems by studying the conservation of energy, momentum, mass, and other quantities \cite{Arnold1989mathematical, Marsden1999introduction, Hairer06}. Thus we consider systems with external forces of arbitrary form. This marks a slight distinction from the port-Hamiltonian formulation, which in addition involves some structure on the external forces, called interaction and control ports in that setting. The term port-Hamiltonian has its origin in control theory, where this formulation is viewed as a special realization of passive systems \cite{van1995hamiltonian, van2006port, van2014port}. There are several recent works on system identification of the port-Hamiltonian formulation of passive systems specifically \cite{Benner2020identification, Cherifi2022non, Morandin2022port}, which take a more classical approach to system identification than we do here, and assume the Hamiltonian to be quadratic.

Among the literature on HNN, several papers have proposed using symplectic integrators during training, since these will preserve the Hamiltonian within some time-independent bound when used for integration \cite{Chen2020Symplectic, Jin2020sympnets, David2021symplectic, Desai2021variational}. It has been argued that using symplectic integrators in the training may allow for extracting qualitative properties e.g.\ through obtaining the Hamiltonian from the so-called inverse modified equation given perfect training \cite{Zhu2020deep, Zhu2022numerical, Offen2022symplectic}. However, convincing theoretical or numerical results on improved accuracy of the resulting models are lacking (see Appendix \ref{sec:integrators}). A study of symplectic integrators for learning pseudo-Hamiltonian systems is to our knowledge non-existing as of yet, but in this paper, we argue that symmetry is a more important property than symplecticity for integrators used for system learning, and is particularly well-suited for dealing with noisy data. We propose to use a fourth-order method, partly motivated by the convincing results from using higher-order methods for neural network models \cite{Desai2021variational, Eidnes2022port, Noren2023learning}.

The works most closely related to ours are the recent papers on system identification of Hamiltonian systems, or a generalization of this, based on approaches similar to SINDy. When applicable, we compare our model to that of \cite{DiPietro2020sparse}, which considers the identification of separable Hamiltonian systems. \cite{Lee2021structure} also considers a port-Hamiltonian formulation, but only in the case where the external force is known a priori. Both these references apply a training algorithm that involves integrating the learned system as proposed for Neural ODEs \cite{Chen2018neural}. By contrast, we use the integration scheme directly in the loss function, as is done for neural network models in \cite{Matsubara2020deep, Jin2020sympnets, David2021symplectic, Eidnes2022port, Noren2023learning}.

\section{Background}
In the experiments presented later in this paper we consider specific Hamiltonian and port-Hamiltonian formulations. However, to demonstrate the broader applicability of our approach, we first introduce a generalized framework, relevant for all systems that can be described by an ODE \eqref{eq:ode}. Therefore we consider a more general formulation of energy-preserving systems than the canonical Hamiltonian formulation
\begin{equation}
    \begin{pmatrix}
        \dot{q} \\
        \dot{p}
    \end{pmatrix}
    =
    \begin{pmatrix}
        0    & I_n \\
        -I_n & 0
    \end{pmatrix}
    \begin{pmatrix}
        \partial H/\partial q \\
        \partial H/\partial p
    \end{pmatrix}
    \quad q, p \in \mathbb{R}^n, H : \mathbb{R}^n \times \mathbb{R}^n \rightarrow \mathbb{R},
\label{canonical_ham}
\end{equation}
considered in e.g.\ \cite{Greydanus2019hamiltonian, DiPietro2020sparse, Lee2021structure}. Furthermore, for non-preserving systems we consider a more general formulation than the port-Hamiltonian formulation of e.g.\ \cite{Benner2020identification, Desai2021port}. Following \cite{Eidnes2022port, Eidnes2023pseudo}, we call this formulation pseudo-Hamiltonian.

\subsection{Hamiltonian systems}
The most general class of energy-preserving systems, encompassing canonical and non-canonical Hamiltonian systems, is given by
\begin{equation}
    \dot{x} = S(x) \, \nabla H(x),
    \label{ham_ode}
\end{equation}
where $x$ denotes the generalized coordinates, $S : \mathbb{R}^d \rightarrow \mathbb{R}^d \times \mathbb{R}^d$ is such that $S(x)$ is an antisymmetric matrix for any $x$, and $H : \mathbb{R}^d \rightarrow \mathbb{R}$ is the Hamiltonian. We continue to call it the Hamiltonian, although it can be any conserved quantity; given initial conditions $x^0$, it will be preserved at all subsequent times, i.e.\ $H(x(t)) = H(x^0)$. This property follows directly from the formulation \eqref{ham_ode} and the antisymmetry of $S$:
\begin{equation}
    \dot{H} = \nabla H^T \dot{x} = \nabla H^T(x) \, S(x) \, \nabla H(x) = 0.
\end{equation}
Even when $H$ is not energy in the physical sense, \eqref{ham_ode} is still usually called an energy-preserving system \cite{Hairer06}. Moreover, an ODE system may have several preserved quantities and can be written on the form \eqref{ham_ode} for different $H$'s and corresponding non-unique $S$'s \cite{McLachlan1999geometric}. If $S$ is the symplectic matrix given in \eqref{canonical_ham}, \eqref{ham_ode} is a canonical Hamiltonian system; if $S(x)$ defines a Poisson bracket satisfying the Jacobi identity, it is a Poisson system \cite{Hairer06}. In all experiments in this paper, we assume that $S$ is constant and known.

\subsection{Pseudo-Hamiltonian formulation}\label{subsec:phf}
We desire a generalization of the Hamiltonian system formulation that allows for energy dissipation and external forces affecting the system. Thus we consider what we call \textit{pseudo-Hamiltonian systems}:
\begin{equation}
    \dot{x} = (S(x)-R(x)) \, \nabla H(x) \, + \, F(x, t), 
\label{port_ham_eq}
\end{equation}
where $R : \mathbb{R}^d \rightarrow \mathbb{R}^d \times \mathbb{R}^d$ is such that $R(x)$ is a positive semi-definite matrix defining the dissipation for each coordinate of $x$, and $F: \mathbb{R}^d \times \mathbb{R} \rightarrow \mathbb{R}^d$ denotes the external forces. If we imposed certain conditions on $F(x,t)$, this would be equivalent with the port-Hamiltonian formulation from control theory \cite{van2006port, van2014port}. The formulation \eqref{port_ham_eq} is completely general and non-unique; e.g., we can set $S(x)=R(x)=0$ and $F(x,t) = g(x,t)$ and recover \eqref{eq:ode}. However, a model based on \eqref{port_ham_eq} sets up a framework for imposing assumptions about the system that will make the formulation less general, as we will demonstrate in Section \ref{sec:experiments}.

\subsection{Numerical integration of ODEs}
When an analytic solution of \eqref{eq:ode} is not available, the evolution of the system from time $t^n$ to time $t^{n+1} =  t^n + \Delta t$ can be estimated by the use of a numerical integration scheme
\begin{equation}\label{eq:scheme}
    \frac{x^{n+1} - x^n}{\Delta t} = \Psi_{\Delta t} (g, x^n, x^{n+1}, t^n),
\end{equation}
where $\Psi$ depends on the chosen integrator. If $\Psi$ is explicitly given by $x^n$ and $t^n$, we say that it is an explicit integrator; we have e.g.\ $\Psi_{\Delta t}(g,x^n,\cdot,t^n) = g(x^n,t^n)$ for the forward Euler method. Implicit integrators depend on $x^{n+1}$ or intermediate steps, and thus a system of equations has to be solved by some root-finding algorithm when these are used for propagating non-linear ODEs. The main drawback of this class of integrators is generally considered to be their computational inefficacy compared to explicit integrators. However, a large class of implicit integrators is explicitly given by $x^n$ and $x^{n+1}$; e.g.\
\begin{equation}\label{eq:midpoint}
\Psi_{\Delta t}(g,x^n,x^{n+1},t^n) = g(\frac{x^n+x^{n+1}}{2},t^n+{\frac{\Delta t}{2}})
\end{equation}
for the implicit midpoint method. These methods are thus well-suited for training models from data where $x^n$ and $x^{n+1}$ are available, as they are then computationally efficient while coming with properties that cannot be achieved with explicit integrators.

\subsection{System Identification}
The system identification approach of \cite{brunton2016sindy} consists of assuming a general form \eqref{eq:ode} and searching for the terms on the right-hand side, thus ideally obtaining the governing equations. That is, given data on $\dot{x}$ and $x$, we assume that $g$ can be expressed by terms from a library of nonlinear functions of $x$, and search for a minimal linear combination $\hat{g}_\theta(x)$ of these that minimizes $\dot{x}-\hat{g}_\theta(x)$. 
In the absence of data on $\dot{x}$, \cite{brunton2016sindy} suggest to approximate this from $x$. In \cite{DiPietro2020sparse, Lee2021structure}, a modified variant of this approach is used to learn the terms of the Hamiltonian and obtain the governing equations by assuming the system has a canonical Hamiltonian formulation. The Hamiltonian function for most conservative dynamical systems will only include a few terms, which means that a small function space may suffice for determining the governing equations. In this paper, we generalize the approach by assuming a pseudo-Hamiltonian formulation, thus also having to learn damping coefficients and external forces. Furthermore, in contrast to \cite{DiPietro2020sparse}, we do not assume separability of the Hamiltonian into potential and kinetic energy, or that the matrix $S$ has the canonical form as in \eqref{canonical_ham}.

A clear advantage of the system identification approach compared to a black-box model such as neural networks is interpretability. Although a neural network may be accurate in prediction given sufficient data and a sufficiently large network \cite{cybenko1989approximation, hornik1991approximation}, it does not reveal the governing equations of the system. System identification models also hold an advantage over black-box models when predicting on data that is drawn from outside the domain of training data; or in other words, they can perform robust data extrapolation \cite{brunton2016sindy}.

\section{Methodology}
Our procedure for system identification is to define function spaces for the Hamiltonian $H$ and possibly the external forces $F$, and train coefficients for each of the terms in the function space while also training coefficients of the damping matrix $R$. Given these search spaces, we will use system identification models to learn the governing equation of dynamical systems with pseudo-Hamiltonian structure. In our implementation, we have included polynomial and trigonometric terms in the function spaces, allowing for a variety of different dynamical systems. 

\subsection{Pseudo-Hamiltonian system identification}
We define the most general PHSI model by
\begin{equation}\label{eq:phsi_full}
    \hat{g}_{\theta}(x, t) := (\hat{S}_\theta(x)-\hat{R}_\theta(x)) \, \nabla \hat{H}_\theta(x) \, + \, \hat{F}_\theta(x, t),
\end{equation}
where $\hat{S}_\theta$, $\hat{R}_\theta$, $\hat{H}_\theta$ and $\hat{F}_\theta$ are each modelling their corresponding term in \eqref{port_ham_eq}. The main advantage of pseudo-Hamiltonian system identification is thus its ability to learn the true equations of a Hamiltonian system that is disturbed by both damping and external forces. Learning the true equations of the inner dynamics of a system can be difficult if it is affected by external forces with a complex form that we have little or no prior knowledge of. Separating out the forces by modeling them with $\hat{F}_\theta$ will make this possible, as we will show in the numerical experiments. $\hat{F}_\theta$ can be modeled either by a system identification model or a neural network, depending on the complexity of the true external forces $F$. We will demonstrate learning $\hat{F}_\theta$ through a system identification model in Section \ref{subsec:msd} and through a neural network in Section \ref{subsec:hybrid}.
In the most general case, the matrices $S(x)$ and $R(x)$ in \eqref{port_ham_eq} would be learned with only the assumption that they are antisymmetric resp.\ positive semi-definite. For the numerical experiments in this paper, we assume to know $S$ exactly and $R$ up to some learnable friction coefficients. 
This reduces the full PHSI model \eqref{eq:phsi_full} to
\begin{equation}
    \hat{g}_{\theta}(x, t) = (S-\mathrm{diag}(\hat{r}_\theta)) \, \nabla \hat{H}_\theta(x) \, + \, \hat{F}_\theta(x, t),
\label{ham_eq}
\end{equation}
where $\mathrm{diag}(\hat{r}_\theta)$ is the matrix with the trainable elements of $\hat{r}_\theta \in \mathbb{R}^d$ on the diagonal. Assuming that the structure matrix $S$ is known gives both an advantage and a disadvantage: On one hand, knowing $S$ allows us to impose the physical law of energy conservation through the Hamiltonian framework. On the other, our approach is restricted to learning models where $S$ is prior knowledge, something that other system identification models such as \cite{brunton2016sindy} are not. However, in the systems we will consider, $S$ is either of canonical form, as is assumed in the many recent works on Hamiltonian neural networks \cite{Greydanus2019hamiltonian}, or a non-canonical form that can be obtained from engineering knowledge. 

\subsection{Pruning} \label{sec:pruning}
We have constructed a simple pruning algorithm for excluding terms during training: For every $P$th epoch, every coefficient whose absolute value has taken on values smaller than a threshold $\epsilon$ for the last $p$ epochs will be set to zero and disregarded for the rest of the training. The pruning thus narrows the search space, making it easier to find the relevant terms and accurate coefficients.  The optimal choices of $\epsilon$ and $p$ are highly problem dependent. In the numerical experiments of Section \ref{sec:experiments}, we have used $p=1$ and found this to work well, and it is this special case that is shown in Algorithm \ref{training_alg}. In Section \ref{sec:pruning_small} we discuss the danger of pruning away terms with small coefficients if $\epsilon$ is not wisely chosen. 

\subsection{Regularization} \label{sec:regularization}
To be able to separate the Hamiltonian system, the damping effects, and the external forces from each other, it is often necessary to implement regularization, especially on the forces. As discussed in Section \ref{subsec:phf}, there is generally not uniqueness in the separation of the pseudo-Hamiltonian system into the internal and external parts. In practice, if $S$ is known and additional structure is imposed on $R$ so that the system would be uniquely defined in the absence of external forces, the model tends to learn the most natural representation, where only the dynamics that do not fit within the Hamiltonian structure with damping are attributed to external forces. We have observed this effect from repeated experiment, and attribute it to it being easier to learn a Hamiltonian $H$ that conforms to the expected structure of a Hamiltonian system than to learn the entirety of dynamics represented by the force function $f$. However, depending on the underlying dynamics and the assumptions imposed on the model, it is often expedient or even necessary to implement regularization, especially on $F$. This is particularly relevant when the external forces are state-dependent and modeled by a neural network, which does not assume any specific form of the term. 

We have used $L_1$-regularization on the coefficients of the Hamiltonian to promote sparsity of the search space, following the SINDy approach \cite{brunton2016sindy}. However, the regularization can negatively affect the accuracy of the trained non-zero parameters. Therefore, we choose to drop the regularization during the second half of training. The idea is that sparsity will be promoted during the first half of training, and then the remaining coefficients will be better tuned during the second half. A study of how regularization affects the structure of the learned PHSI model is provided in Appendix \ref{appendix_regularization}. 

\subsection{Loss function} \label{sec:loss}
Several earlier works on system identification and HNN have assumed that the true derivatives of the system are known or can be approximated by e.g.\ finite differences \cite{brunton2016sindy, Greydanus2019hamiltonian, Desai2021port}. Other works rely on an explicit integrator to propagate the learned system and evaluate the loss by comparing to available data on state variables \cite{Chen2020Symplectic, Desai2021variational, DiPietro2020sparse}. In this paper, we train on an integration scheme \eqref{eq:scheme}, which to our knowledge has not previously been proposed for system identification. The main advantage is that we can use data from two successive points, thus partly averaging out the noise at no extra computational cost. We propose to use symmetric non-partitioned integrators, which depend equally on $x^n$ and $x^{n+1}$ when evaluating $\Psi_{\Delta t}$ in \eqref{eq:integrationscheme}. E.g., if the data points $x^n$, $n=1,\ldots,N$, have independent Gaussian noise with standard deviation $\sigma$, the standard deviation of the noise of $(x^n+x^{n+1})/2$ in \eqref{eq:midpoint} is $\sigma/\sqrt{2}$. The midpoint method is however only a second-order integrator.
For all experiments, we train the PHSI models using the fourth-order symmetric integration scheme introduced in \cite{Eidnes2022port}.
Appendix \ref{sec:integrators} includes the definition of this integrator and a discussion of properties that can influence the choice of the integrator, like symmetry, symplecticity, and order.

When the numerical integrator $\Psi$ has been chosen, the loss function is evaluated on the corresponding integration scheme:
\begin{equation}
\begin{split}
    L = & \, \lVert \frac{x^{n+1} - x^n}{\Delta t} - \Psi_{\Delta t}(\hat{g}_\theta, x^n, x^{n+1}, t^n) \rVert _2^2 \\
    & \, + \lambda_H \lVert \theta_H \rVert_1 + \lambda_F \lVert \theta_F \rVert_1 + \lambda_R \lVert \theta_R \rVert_1,
\end{split}
\label{loss}
\end{equation}
given for one data point $x^n$. Here $\lambda_H$, $\lambda_F$, and $\lambda_R$ are regularization parameters, and $\theta_H$, $\theta_F$ and $\theta_R$ denote the vectors of trainable parameters in $\hat{H}_\theta$, $\hat{F}_\theta$ and $\hat{R}_\theta$. If $\hat{F}$ is modelled by a neural network and not a system identification model, the term $\lambda_F \lVert \theta_F \rVert_1$ is replaced by $\lambda_F \lVert \hat{F}_\theta(\frac{x^n+x^{n+1}}{2}) \rVert_1$. As noted above, $\lambda_H$, $\lambda_F$, and $\lambda_R$ may be set to zero midway during training. Pseudocode for how the loss is computed and how the model is trained is shown in Algorithm \ref{training_alg}.

\begin{algorithm}[tb]
    \caption{Training algorithm}
    \begin{algorithmic}
        \STATE Input variables: input data $X = [x_1,..., x_m]$, number of epochs $E$, integrator $\Psi$, learning rate $\eta$, batch size $b$, pruning interval $P$, pruning threshold $\varepsilon$
        \FOR{$e$ in $(1, ..., E)$}
            \STATE Create a new permutation of $p = \{p_1, ..., p_m\} \text{of} \{1, ..., m\}$
            \FOR{$i$ in $(1, ..., m/b)$}
                \STATE Select $b$ samples from the shuffled dataset $X_{\text{batch}} = \{ [x{p_j}] \}_{j=ib}^{(i+1)b}$
                \STATE Compute gradients of the loss with respect to the model parameters through backpropagation: \begin{itemize}
                    \item $G_{\Xi_H} = \frac{dL(\hat{g}_{\text{PHSI}}; X_{\text{batch}}, \Psi)}{d \Xi_H}$
                    \item $G_{\Xi_F} = \frac{dL(\hat{g}_{\text{PHSI}}; X_{\text{batch}}, \Psi)}{d \Xi_F}$
                    \item $G_{\hat{R}} = \frac{dL(\hat{g}_{\text{PHSI}}; X_{\text{batch}}, \Psi)}{d \hat{R}}$
                \end{itemize}
                \STATE Update model parameters: $\Xi_H, \Xi_F, \hat{R} \leftarrow (\Xi_H, \Xi_F, \hat{R}) - \eta(G_{\Xi_H}, G_{\Xi_F}, G_{\hat{R}})$
            \ENDFOR
            \IF{$e//P = 0$}
                \FOR{$\xi$ in $\{ \Xi_H, \Xi_F, \hat{R} \}$}
                    \IF{$|\xi| < \varepsilon$}
                        \STATE $\xi=0$ for the rest of the training
                    \ENDIF
                \ENDFOR
            \ENDIF
        \ENDFOR
    \end{algorithmic}
\label{training_alg}
\end{algorithm}

\subsection{Initializations and hyperparameters}
When initializing the model, all coefficient parameters are set to $0.2$. We want to avoid initial values of the parameters near zero, as this increases the risk of terms being removed by the pruning algorithm too early. We use the Adam optimizer and have to specify the learning rate, batch size and number of epochs to run. The other hyperparameters that have to be specified are parameters for regularization on the external forces and the damping, and the pruning parameters $P$ and $p$. These are given for the specific experiments in Section \ref{sec:experiments}.


\subsection{Baseline system identification model} \label{baseline}
To evaluate the effect of imposing a pseudo-Hamiltonian structure on our model, we have tested the PHSI model against a baseline system identification (BSI) model trained with a similar strategy, including the same integrator, but without assuming a pseudo-Hamiltonian structure. This model learns a sparse representation of $g$ in \eqref{eq:ode} from a function space similar to that used for learning $H$ in the PHSI model, albeit with a constant term and one degree lower polynomials to reduce the search space. If we for instance want to learn a system with a Hamiltonian of polynomial terms of degree $n$, the search space for system identification of the Hamiltonian has ${d+n\choose n} -1$ terms. A system identification model directly learning a representation of $g$ will have a search space of $d \cdot {d+n-1\choose n-1}$, which is more than $\frac{dn}{d+n}$ times as many terms. Since this term can grow large for large values of $d$ and $n$, the search space of $g$ can be much larger than that of the Hamiltonian $H$, making it more difficult to learn through system identification. This difference in structure can give the BSI model a disadvantage compared to the PHSI model.

\section{Experiments}\label{sec:experiments}
To assess the performance of the PHSI model, we compare it against three other models: the SINDy model of \cite{brunton2016sindy}, learning $g$ in (\ref{eq:ode}); the sparse symplectically integrated neural networks (SSINN) of \cite{DiPietro2020sparse}, which learns a separable Hamiltonian, assuming the structure in (\ref{canonical_ham}); and the baseline model (BSI) presented in Section \ref{baseline}, learning $g$ in (\ref{eq:ode}). 
For each test, we train the relevant models on two data sets: one noise-free and one with Gaussian noise: $x^{i}_{\text{noise}} = x^{i} + \epsilon^i$ for all $x^{i}$, where $\epsilon^i \sim N(0, \sigma^2)$. When choosing the polynomial degree of the search space, SINDy and BSI will use one polynomial degree lower than PHSI, since they train on the derivative terms of the Hamiltonian. The data sets consist of trajectories with random initialization.

The Adam optimizer is used with a weight decay constant of $10^{-4}$. The learning rate is chosen from the following search space: $\{10^{-2}, 3\cdot10^{-3}, 10^{-3}\}$. The training data is shuffled for all experiments, and the integrator is the fourth-order symmetric integration scheme introduced in \cite{Eidnes2022port}; see Appendix \ref{sec:integrators} for details on this. The batch size is 32. For the PHSI and BSI models, polynomial coefficients have an initial value of $0.2$, and trigonometric coefficients (amplitude and frequency) have an initial value of $1$.

\subsection{Learning a separable Hamiltonian system} \label{subsec:hh}
We first test our methodology on data obtained from a pure Hamiltonian system. Thus we can benchmark against existing methods while demonstrating how our proposed framework facilitates assumptions being imposed. We consider the H\'enon--Heiles system for describing the two-dimensional chaotic motion of stars around a galactic center \cite{Henon1964applicability}. It is a canonical system of the form (\ref{canonical_ham}) with $n=2$, and the Hamiltonian is defined as
\begin{equation*}
    H(q, p) = \frac{1}{2} (q_1^2 + q_2^2 + p_1^2 + p_2^2) + q_1^2 q_2 - \frac{1}{3} q_2^3.
\end{equation*}
Because this is a separable  Hamiltonian system, it can be modeled by SSINN.

To ensure a fair comparison, we train on the data generated and used in \cite{DiPietro2020sparse}, although with more noise. 
The data consists of $3000$ pairs of $q$, $p$ at $t=0$ and $t=0.1$. The noisy data have $\sigma = 0.02$, which corresponds to approximately $3\%$ of the maximum absolute values in the data. For the results reported here, the learning rate is $3\cdot10^{-3}$, and regularization is not used. The PHSI and BSI models are each trained for 60 epochs. The $P$-value and $\epsilon$-value in the pruning algorithm (\ref{training_alg}) are set to $5$ and $0.05$, respectively. The search space for PHSI and SSINN is third-degree polynomials, while for SINDy and BSI it is second-degree polynomials.

\begin{figure}[tb]
    \centering
    \includegraphics[scale=0.6]{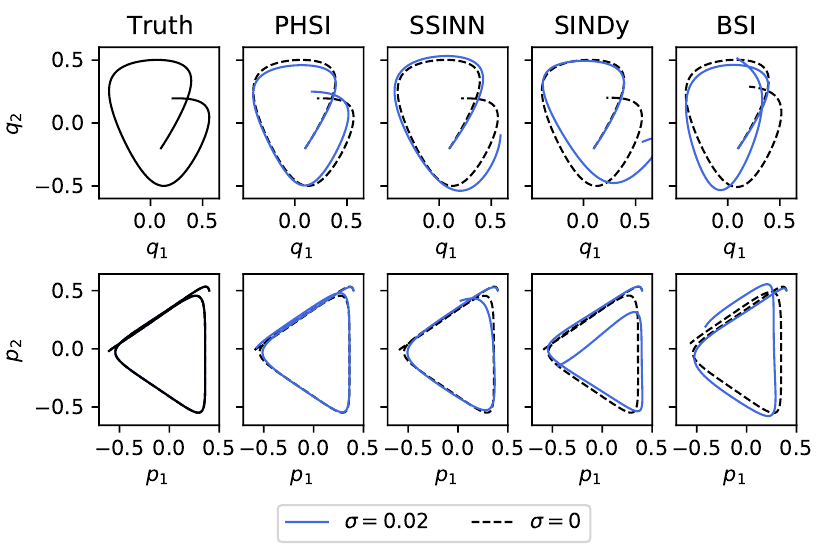}
    \caption{Comparison of simulated trajectories for the H\'enon--Heiles system, with initial value $(q,p) = (0.1, -0.2, 0.4, 0.5)$, from $t=0$ to  $t=10$.}
    \label{hh_trajectory}
\end{figure}

\begin{figure}[ht]
    \centering
    \includegraphics[scale=0.6]{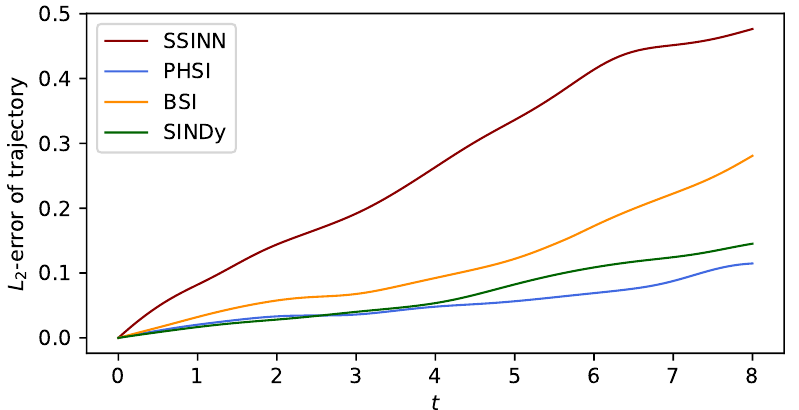}
    \caption{The average $L_2$-error of the trajectories obtained with 10 different random initial conditions, from the different models of the H\'enon--Heiles system trained on noisy data with $\sigma = 0.02$, compared to trajectories simulated from the exact system, }
    \label{hh_l2error}
\end{figure}

Figure \ref{hh_trajectory} shows example trajectories obtained from integrating the different learned systems in time. When training on noise-free data, all models learn the true coefficients up to $10^{-2}$, while noise affects the different methods to different degrees. On the noisy data, PHSI outperforms SSINN by one order of magnitude when comparing the $L_2$-error of the trajectories, as depicted in Figure \ref{hh_l2error}. PHSI and BSI both perform better than SSINN and SINDy, suggesting that the fourth-order symmetric integrator handles the noise well. The superior performance of PHSI over BSI suggests that imposing the Hamiltonian structure gives better performance yet. 
The SINDy and BSI models could not be converted to Hamiltonian functions when trained on noisy data, and the solution of the SINDy model tends to become unstable over long time periods. Table \ref{hh_table} compares the coefficients learned from training on the noisy data set for the two models imposing a Hamiltonian structure.
\begin{table}[ht]
\centering
\caption{Learned coefficients in the Hamiltonian for the H\'enon--Heiles system.}
\label{hh_table}
\begin{tabular}{clcccccccc}
\toprule
& & $q_1^2$ & $q_2^2$ & $q_1^2 q_2$ & $q_2^3$ & $p_1^2$ & $p_2^2$ \\
\midrule
True Value & & $0.5$ & $0.5$ & $1$ & $-0.333$ & $0.5$ & $0.5$ \\
PHSI & & $0.509$ & $0.495$ & $1.009$ & $-0.338$ & $0.501$ & $0.477$ \\
SSINN & & $0.421$ & $0.378$ & $0.569$ & $-0.195$ & $0.484$ & $0.443$ \\
\midrule
& & $q_2 p_1^2$ & $q_2$ & $q_1q_2$ & $p_2$ & $p_1p_2$ \\
\midrule
True Value & & $0$ & $0$ & $0$ & $0$ & $0$ \\
PHSI & & $0.124$ & $0$ & $0$ & $0$ & $0$ \\
SSINN & & N/A & $0.004$ & $0.002$ & $-0.006$ & $-0.006$ \\
\bottomrule
\end{tabular}
\end{table}

\subsection{Learning a non-separable Hamiltonian system} \label{subsec:nls}
A strength of PHSI is that it can learn non-separable Hamiltonians as well. We demonstrate this by testing its ability to learn a finite-dimensional nonlinear Schrödinger system, as considered in \cite{tao2016explicit}. This is a canonical system \eqref{ham_ode} with $d=2$ and the Hamiltonian
\begin{align*}
    H(q,p) = & \frac{1}{4} (q_1^2 + p_1^2)^2 + \frac{1}{4}(q_2^2 + p_2^2)^2 \\
    & \, -q_1^2 q_2^2 - p_1^2 p_2^2 + q_1^2 p_2^2 + q_2^2 p_1^2 - 4 q_1 q_2 p_1 p_2.
\end{align*}

We use two sets of training data: one clean and one noisy with $\sigma = 5 \cdot 10^{-4}$. The training sets consist of $30$ trajectories consisting of $100$ points with step size $0.01$, each randomly initialized from a uniform distribution between $0$ and $1$. We have trained the PHSI, SINDy and BSI models, and compare the performances of these. For the PHSI model, the learning rate is $0.01$, and regularization is not used. The model is trained for 100 epochs. We use $P=20$ and $\epsilon=0.05$ in the pruning algorithm. The search space for PHSI is fourth-degree polynomials, while for SINDy and BSI it is third-degree polynomials.

\begin{figure}[ht]
    \centering
    \includegraphics[scale=.71]{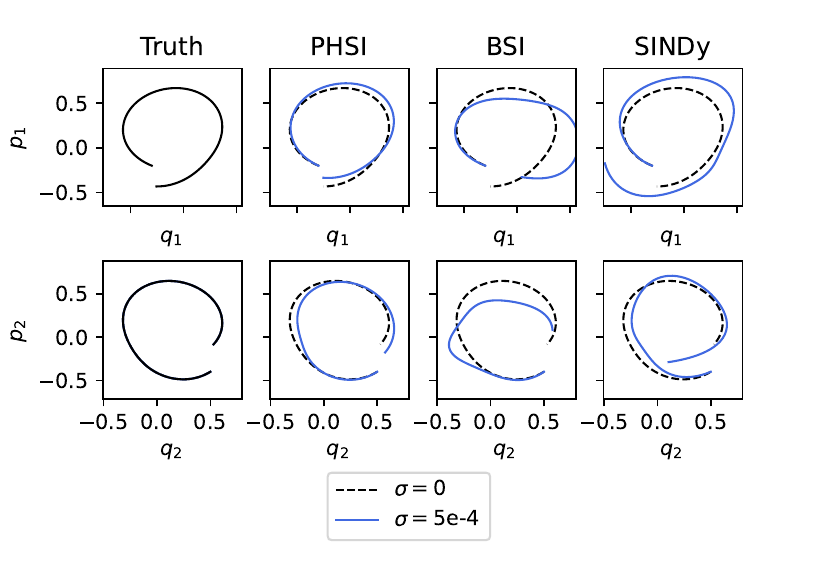}
    \caption{
    Phase portraits showing the trained models' trajectories next to the ground truth trajectory, for the nonlinear Schrödinger system. All models are trained on the two training sets, one clean data set and one noisy with $\sigma= 10^{-4}$. The initial values are $(q,p) = (-0.3, 0.5, -0.2, -0.4)$.}
\end{figure}
\begin{figure}[ht]
    \centering
    \includegraphics[scale=.71]{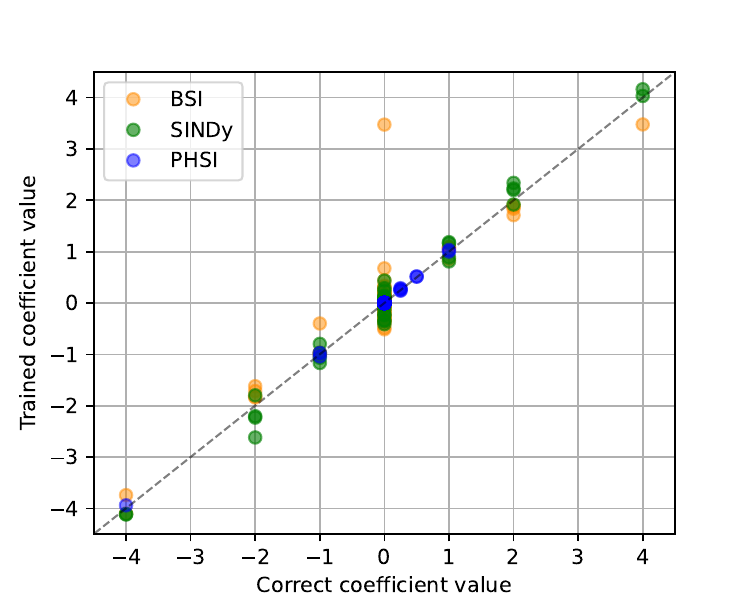}
    \caption{
    The trained coefficients of the PHSI, BSI, and SINDy models of the nonlinear Schrödinger system, plotted against the respective true values. A perfectly trained model will only have points along the dotted line. The models are trained on the noisy data set. Note that PHSI learns the Hamiltonian function while SINDy learns the right-hand side $g$ of \eqref{eq:ode} and hence they will not learn the same coefficients for corresponding terms. 
    }
    \label{nls_coeff}
\end{figure}

PHSI and SINDy are able to learn the true equations of the system on the noise-free data up to a precision of $10^{-5}$, while BSI is able to learn to a precision of $10^{-3}$. On the noisy data, PHSI learns the true equations up to a precision of $10^{-1}$ and excluding $53$ out of $58$ terms correctly by the pruning algorithm. SINDy only learns the terms up to a precision of $1$, and BSI does not even achieve this precision and also exclude many terms incorrectly during pruning. Figure \ref{nls_coeff} confirms a more accurate representation of the true system equations in the PHSI model than SINDy when trained on noisy data. Neither the coefficients learned by SINDy nor BSI could be converted to a separable Hamiltonian. 

\subsection{Learning a pseudo-Hamiltonian system}\label{subsec:msd}
Consider a mass-spring system with damping and external forces. 
The Hamiltonian is given by
\begin{equation*}
    H(q, p) = \frac{k}{2} q^2 + \frac{1}{2m} p^2,
\end{equation*}
where $k$ is the stiffness constant of the spring and $m$ is the mass. Furthermore, we have damping and an external force affecting the momentum, so that a pseudo-Hamiltonian formulation \eqref{port_ham_eq} of the system is given by
\begin{equation}
    \begin{bmatrix}
        \dot{q} \\
        \dot{p}
    \end{bmatrix}
    = 
    \begin{bmatrix}
        0 & 1 \\
        -1 & -c
    \end{bmatrix}
    \begin{bmatrix}
        k q \\ \frac{1}{m}p
    \end{bmatrix}
    + 
    \begin{bmatrix}
        0 \\ \alpha \sin{(\omega t)}
    \end{bmatrix}.
    \label{msd_system}
\end{equation}
We let $k=m=1$, $\alpha=2$, $\omega=0.5$ and $c=0.3$. The data consists of 50 trajectories from time $0$ to $10$, with time step $0.1$, and the noisy data set has $\sigma = 0.2$. 

We train three system identification models, PHSI, BSI, and SINDy, and also a PHNN model as described in \cite{Desai2021port}. For the PHSI model, we train a third-degree polynomial function space for the Hamiltonian, while the function space for the external force is a third-degree polynomial as well as trigonometric functions where both amplitude and frequency are trainable parameters, and the force is assumed to be strictly time-dependent and only directly affecting the momentum. In addition, we assume that damping only directly affects the momentum, so that learning the matrix $R$ in this case reduces to learning the friction coefficient $c$. Both the BSI and SINDy models are trained on a third-degree polynomial (including a constant) plus trigonometric functions.
The SINDy model has a disadvantage as it can only learn autonomous systems. To still be able to use it, we convert the system into an autonomous one by adding time as a variable to (\ref{msd_system}): $\dot{t} = 1$. The PHSI, BSI, and PHNN models are run for 150 epochs, with a learning rate $5\cdot  10^{-3}$. We set $P = 20$ and $\epsilon=0.05$ in the pruning algorithm. $L_1$-regularization is used for the PHSI model on the forces and the Hamiltonian function: $\lambda_H=0.1$ and $\lambda_F=0.01$ in (\ref{loss}). The external force of the PHNN model has a regularization parameter of $0.1$.

\begin{figure}[ht]
    \centering
    \includegraphics[scale=0.6]{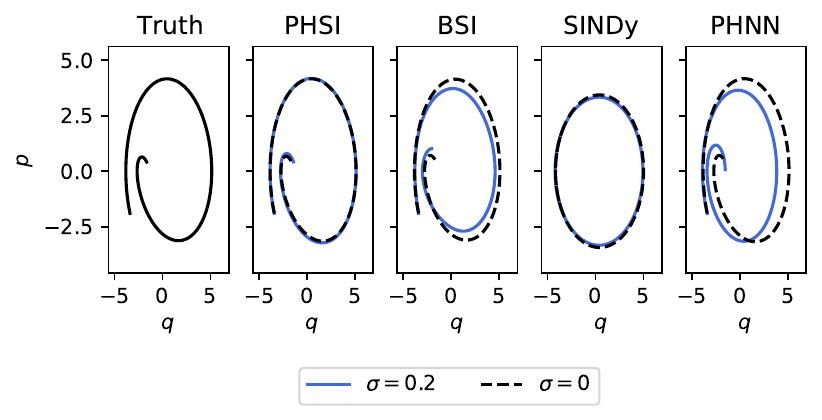}
    \caption{Comparison between the phase portraits obtained from integrating the exact forced and damped mass-spring system and the learned models from time $0$ to $10$. The initial value is $(q,p) = (-3.4, -1.9)$.}
    \label{msd_traj}
\end{figure}

\begin{figure}[ht]
    \centering
    \includegraphics[scale=0.6]{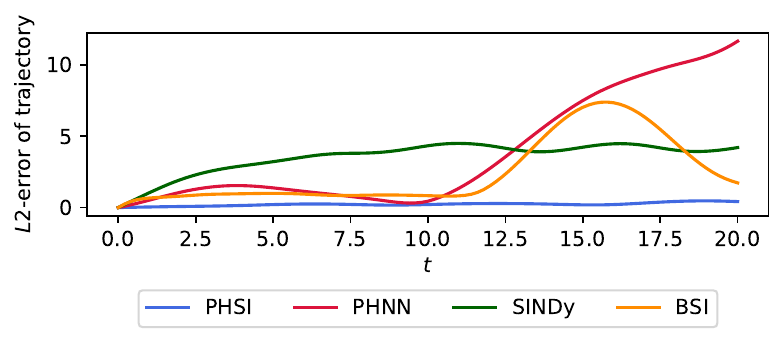}
    \caption{Average $L_2$-error of 30 simulated trajectories of the forced and damped mass-spring system, with random initial conditions.}
    \label{msd_error_noise}
\end{figure}

Predictions resulting from the models are shown in Figure \ref{msd_traj} and coefficients are given in Table \ref{msd_table}. 
From Figure \ref{msd_error_noise} we see that PHSI clearly outperforms PHNN on predictions beyond the time period of the training data, which is expected since the external force is time-dependent. 
Even on noisy data, the PHSI model is able to separate the inner dynamics, the damping effects and the external forces from each other, while the BSI model did not learn terms that can be separated into inner and outer dynamics in any obvious way. This provides a qualitative argument for using a model that assumes a pseudo-Hamiltonian formulation.

\begin{table}[ht]
\centering
\caption{Learned coefficients for the forced and damped mass-spring problem. $q$ and $p$ are multiplied with the trained coefficients while $c, \alpha, \omega$ and $const.$ are themselves the trainable parameters. Empty entries mean that the model does not learn that term.}
\label{msd_table}
\begin{tabular}{clccccc}
\toprule
\multicolumn{7}{c}{Trained PHSI parameters} \\
\midrule
 & $q^2$ & $p^2$ & $c$ & $\alpha$ & $\omega$ \\
\midrule
True value & $0.5$ & $0.5$ & $0.3$ & $2$ & $0.5$ \\
$\hat{H}$ & $0.473$ & $0.484$ & & & & \\
$\hat{R}$ & & & 0.300 & & & \\
$\hat{F}$ & 0 & 0 & 0 & 1.984 & 0.505 \\
\midrule
\multicolumn{7}{c}{Trained BSI and SINDy parameters} \\
\midrule
& $q$ & $p$ & $const.$ & $\alpha$ & $\omega$ \\
\midrule
True $\hat{\dot{q}}$ & 0 & 1 & 0 & 0 & 0\\
BSI $\hat{\dot{q}}$  & 0 & 0.991 & 0 & 0 & 0 \\
SINDy $\hat{\dot{q}}$  & 0 & 0.983 & 0 & 0 & 0 \\
\midrule
True $\hat{\dot{p}}$ & $-1$ & $-0.3$ & $0$ & $2$ & $0.5$ \\
BSI  $\hat{\dot{p}}$  & $-0.980$ & $-0.278$ & $0$ & $1.999$ & $0.506$ \\
SINDy $\hat{\dot{p}}$ & $-0.548$ & $0$ & $0.241$ & & & \\
\bottomrule
\end{tabular}
\end{table}

\subsection{Hybrid model combining system identification with a neural network}\label{subsec:hybrid}
In reality, many systems will be affected by external forces for which it can be difficult to find analytic terms. The final system to be studied consists of $N$ tanks connected by $M$ pipes, also considered in \cite{Eidnes2022port}, with leaks that can be viewed as external forces. The incidence matrix $B \in \mathbb{R}^{M \times N}$ describes how the pipes and tanks are connected. The friction in the pipes is assumed to depend linearly on the flow, and the total energy is given by
\begin{align}
    H(\phi, \mu) = \sum_i^M \frac{1}{2J_i} \phi_i^2 + \sum_j^N \frac{g\rho}{2 A_j} \mu_j^2,
\label{tank_hamiltonian}
\end{align}
where $\phi_i$ is the flow in pipe $i$ scaled by a factor $J_i$ depending on the density of the fluid and the dimension of the pipe, $\mu_j$ is the volume of the fluid in tank $j$, $g$ is the gravitational constant, $\rho$ is the density of the fluid, and $A_j$ is the footprint of tank $j$.
We thus have the pseudo-Hamiltonian formulation
\begin{align}
    \begin{bmatrix}
        \dot{\phi} \\
        \dot{\mu} 
    \end{bmatrix}
    = 
    \begin{bmatrix}
        -\text{diag}(r_p) & B^T \\
        B & 0_{N\times N} 
    \end{bmatrix}
    \begin{bmatrix}
        \frac{\partial H}{\partial \phi} \\
        \frac{\partial H}{\partial \mu}
    \end{bmatrix}
    +
    \begin{bmatrix}
        0_M \\
        f(\phi,  \mu)
    \end{bmatrix},
\end{align}
where $r_p \in \mathbb{R}^M$ contains the friction coefficients relating to each pipe. When simulating the system to generate the training data, we set $\rho=1$, $J_i = 0.02$  $\forall i$, $A_j =1$  $\forall j$, $r_p =  (0.03, 0.03, 0.09, 0.05, 0.05)$. A leak in the fourth tank is described by $f(\phi, \mu) = (0,0,0,-10\min{(0.3, \max{(\mu_4, 0.3)})})$. Initial conditions are uniformly drawn, $-1 < x^i <1 $, for all $i$. 

We compare SINDy and BSI against a hybrid PHSI model that only aims to find analytic terms for the internal dynamics. It assumes that the external force only affects the fourth tank, and models this by a neural network. This consists of three hidden layers, 100 nodes per layer, and the ReLU activation function in each layer. 
The training data consists of 60 trajectories from $t= 0$ to $t=0.5$, with time step $0.01$.
The PHSI and BSI models are trained for 100 epochs, with learning rate $3 \cdot 10^{-2}$ and $P=10$ and $\epsilon=0.05$. For the PHSI model, $L_1$ regularization is used on the external forces and the Hamiltonian with $\lambda_H=0.5, \lambda_F=0.001$ in (\ref{loss}). The search space for $\hat{H}_\theta$ is second-degree polynomials. The BSI and SINDy models has a search space of first-degree polynomials.

\begin{figure}[ht]
    \centering
    \includegraphics[scale=0.6]{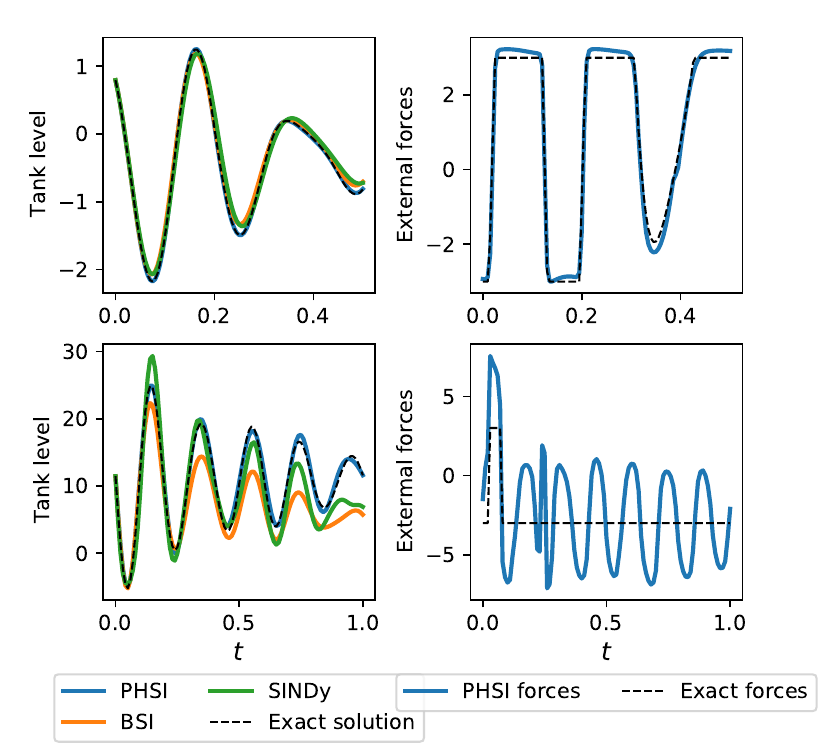}
    \caption{Simulations of the tank system. \textit{Left:} The volume of the fluid in the leaky fourth tank simulated from the exact system and the different models. \textit{Right:} The leak approximated by the neural network in the hybrid PHSI model, compared to the exact solution. The upper plots have initial conditions within the distribution of the training data: $(\phi,\mu)= (-0.4, 0, 0.5, 0, 0.2, 0 -0.6, -0.5, 0.5)$. The lower plots show extrapolation in time and space; time from $0$ to $1$ and initial state values $(\phi,\mu)= (10, 19, 4, 19, 7, 9, 17, 9, 11)$.}
    \label{tank_trajectory_extr}
\end{figure}

\begin{figure}[ht]
    \centering
    \includegraphics[scale=0.6]{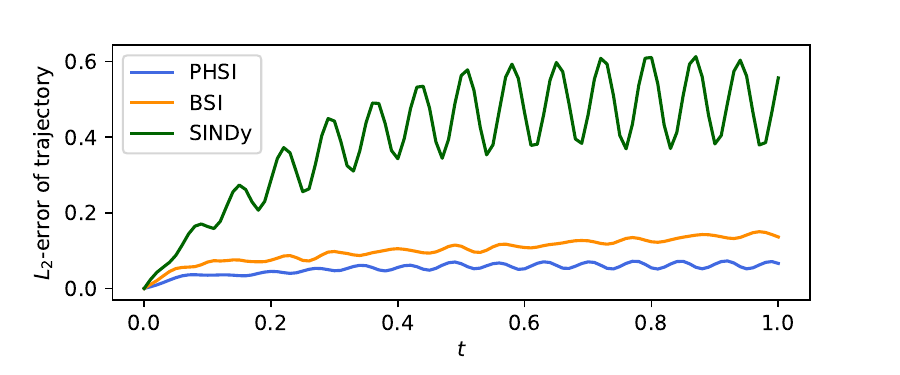}
    \caption{Average $L_2$-error of 30 sets of simulated future tank volumes and pipe flows in the tank system, trained on noisy data with $\sigma=0.005$.}
    \label{tank_error}
\end{figure}

Figure \ref{tank_trajectory_extr} provides example trajectories obtained from the models and Figure \ref{tank_error} shows the error averaged over predictions from 30 random initial conditions. The PHSI model learned the true Hamiltonian up to $10^{-1}$ precision, as well as learning a precise neural network model of the leak. The learned PHSI parameters are shown in Table \ref{fig:tank_table}.
BSI and SINDy did not learn terms that could be converted to a Hamiltonian function and have no clear separation of the internal system from the external forces. This indicates that they are not able to accurately learn the true system, and instead learn polynomial approximations that approximate the system well only within the domain of the training data. 
The lower plots in Figure \ref{tank_trajectory_extr} confirm this. When testing the learned models on initial conditions well outside the range of the training data, PHSI greatly outperforms the other models. 
\begin{table}[ht]
\centering
\caption{Learned coefficients for the tank system on the noisy data. $x_1^2, \dots, x_9^2$ are multiplied with the trained coefficients while $r_1, \dots, r_5$ are themselves the trainable parameters.}
\label{fig:tank_table}
\begin{tabular}{clcccccc}
\toprule
& $x_1^2$ & $x_2^2$ & $x_3^2$ & $x_4^2$ & $x_5^2$ & $x_6^2$ & $x_7^2$ \\
\midrule
True value & $25$ & $25$ & $25$ & $25$ & $25$ & $4.905$ & $4.905$ \\
PHSI & $24.94$ & $24.98$ & $24.98$ & $24.95$ & $24.97$ & $4.930$ & $4.960$ \\
\midrule
& $x_8^2$ & $x_9^2$ & $r_1$ & $r_2$ & $r_3$ & $r_4$ & $r_5$ \\
\midrule
True value & $4.905$ & $4.905$ & $0.03$ & $0.03$ & $0.09$ & $0.05$ & $0.05$ \\
PHSI & $4.890$ & $4.930$ & $0.031$ & $0.029$ & $0.086$ & $0.051$ & $0.041$ \\
\bottomrule
\end{tabular}
\end{table}

\section{Pruning and small coefficients}\label{sec:pruning_small}
The pruning technique outlined in Section \ref{sec:pruning} and Algorithm \ref{training_alg} introduces the hyperparameters $P$, $\epsilon$ and $p$. These parameters can greatly affect the training, and they should therefore be set with care and depending on the problem. In particular, no coefficent smaller than $\epsilon$ can be learned, so it should not be set too high. On the other hand, the pruning becomes more challenging with lower values of $\epsilon$.

Thus it is clear that the PHSI method, like other system identification methods, will struggle to learn terms with small coefficients. To test this, we apply the same methods as in Section \ref{subsec:msd} on a slightly skewed mass-spring system with the same damping and external force. That is, we modify the Hamiltonian to be
\begin{equation}\label{eq:Hmod}
    H(q, p) = \frac{1}{2} q^2 + \frac{1}{2} p^2 + \gamma q p,
\end{equation}
for $\gamma = \{0.1, 0.07, 0.05, 0.03, 0.01, 0.007, 0.005, 0.003, 0.001\}$ and train and test the methods on the resulting system. We try $\epsilon = \{0.05, 0.01, 0.001\}$, and use otherwise the same hyperparameters for the pruning as in Section \ref{subsec:msd}, where $\epsilon = 0.05$ and $\gamma=0$. When $\epsilon=0.05$ or $\epsilon=0.01$, the PHSI method prunes away all terms that should be zero, whether it trains on exact or noisy data, for all choices of $\gamma$. However, when $\epsilon=0.001$, it fails to prune away some of the terms that should be zero when training on the noisy data. E.g., when $\gamma=0.003$ we get
\begin{align*}
    &\hat{H}_\theta(q,p) = 0.4987 q^2 + 0.5004 p^2 + 0.0028 qp,\\
    &\hat{c}=0.2978,\\
    &\hat{f}_\theta(t) = 1.9767 \sin{(0.4993 t)}
\end{align*}
on the noise-free data, but
\begin{align*}
    &\hat{H}_\theta(q,p) = 0.0044 p^3 + 0.4845 q^2 + 0.5075 p^2 - 0.0196 qp + 0.0012 q - 0.0098 p,\\
    &\hat{c}=0.2709,\\
    &\hat{f}_\theta(t) = 1.9334 \sin{(0.4951 t)}
\end{align*}
on the noisy data.

As we see in Table \ref{tab:pruningtest}, the PHSI model learns a non-zero approximation of $\gamma$ when it trains on noise-free data and $\gamma > \epsilon$, with the sole exception $\gamma=0.005, \epsilon=0.001$. When the data is noisy, we see in Table \ref{tab:pruningtest_noisy} that the method prunes away the $qp$ term when $\gamma=0.07, \epsilon=0.05$ and $\gamma=0.005, \epsilon=0.001$, but otherwise learns non-zero approximations of $\gamma>\epsilon$.

\begin{table}[ht]
\centering
\caption{The learned approximations of $\gamma$ in \eqref{eq:Hmod}, for different $\gamma$ and different pruning tresholds $\epsilon$, trained on noise-free data.}
\label{tab:pruningtest}
\resizebox{\textwidth}{!}{
\begin{tabular}{c|ccccccccc}
\toprule
$\epsilon \backslash \gamma$ & $0.1$ & $0.07$ & $0.05$ & $0.03$ & $0.01$ & $0.007$ & $0.005$ & $0.003$ & $0.001$ \\
\midrule
$0.05$ & $0.1012$ & $0.0718$ & $0$ & $0$ & $0$ & $0$ & $0$ & $0$ & $0$ \\
$0.01$ & $0.1000$ & $0.0717$ & $0.0501$ & $0.0310$ & $0$ & $0$ & $0$ & $0$ & $0$ \\
$0.001$ & $0.1006$ & $0.0711$ & $0.0489$ & $0.0301$ & $0.0102$ & $0.0051$ & $0$ & $0.0028$ & $0$ \\
\bottomrule
\end{tabular}
}
\end{table}

\begin{table}[ht]
\centering
\caption{The learned approximations of $\gamma$ in \eqref{eq:Hmod}, for different $\gamma$ and different pruning tresholds $\epsilon$, trained on data with added Gaussian noise with standard deviation $\sigma=0.2$.}
\label{tab:pruningtest_noisy}
\resizebox{\textwidth}{!}{
\begin{tabular}{c|ccccccccc}
\toprule
$\epsilon \backslash \gamma$ & $0.1$ & $0.07$ & $0.05$ & $0.03$ & $0.01$ & $0.007$ & $0.005$ & $0.003$ & $0.001$ \\
\midrule
$0.05$ & $0.0974$ & $0$ & $0$ & $0$ & $0$ & $0$ & $0$ & $0$ & $0$ \\
$0.01$ & $0.1001$ & $0.0485$ & $0.0515$ & $0.0412$ & $0$ & $0$ & $0$ & $0$ & $0$ \\
$0.001$ & $0.0833$ & $0.0754$ & $0.0372$ & $0.0399$ & $0.0034$ & $0.0093$ & $0$ & $-0.0196$ & $-0.0062$ \\
\bottomrule
\end{tabular}
}
\end{table}
\section{Conclusion}
This paper presents several advances in the identification of physical systems that can be modeled as ODEs. For one, we demonstrate improved learning of Hamiltonian systems, compared to standard methods and a recent model specifically developed for such systems. Secondly, we develop the methodology further to be applicable to the much wider class of pseudo-Hamiltonian systems. Lastly, we show how we can incorporate neural network models in our method to separate out external forces that are difficult to identify by analytic terms. In all experiments, the PHSI model performs especially well on noisy data. We attribute this both to the model assumption of an internal Hamiltonian structure and a training strategy that automatically averages out some of the noise.

We have not emphasized computational cost in our numerical examples, and have not yet made a significant effort to optimize our implementation of PHSI or BSI. Compared to SINDy, our models use significantly longer time to train to convergence. This will be addressed in future research, as will extensions to more complex systems, including partial differential equations.

\section*{Acknowledgments}
The authors thank Vegard Antun and Alexander Stasik for fruitful discussions and insightful comments.

\bibliography{phsi}
\bibliographystyle{abbrv}


\newpage
\appendix

\section{On numerical integration schemes for learning dynamical system models}\label{sec:integrators}

Consider a known ODE \eqref{eq:ode}. A numerical integration scheme
\begin{equation}\label{eq:integrationscheme}
\frac{x^{n+1}-x^n}{\Delta t} = \Psi_{\Delta t}(g,x^n,x^{n+1},t^n)
\end{equation}
can be used to find the approximate solution $x^{n+1} \approx x(t+\Delta t)$ given $x^n \approx x(t)$. Such a scheme is \textit{implicit} if $\Psi_{\Delta t}$ depends on $x^{n+1}$ or intermediate steps and \textit{explicit} otherwise. E.g., we have the explicit (forward) Euler method by $\Psi_{\Delta t}(g,x^n,x^{n+1},t^n) = g(x^n,t^n)$ and the implicit (backward) Euler method by $\Psi_{\Delta t}(g,x^n,x^{n+1},t^n) = g(x^{n+1},t^n+\Delta t)$.

The main drawback of using an implicit integrator for numerical integration is computational cost. An implicit scheme generally results in a system of equations that have to be solved using a root-finding algorithm like Newton's method. However, when we have available successive data points, other considerations should be taken, and we argue that we should use the available data to alleviate the effect of noise. Thus we suggest to use an integration scheme that relies equally on the two successive data points, averaging out the noise from two measurements. To that end, we consider a training strategy similar to that of \cite{Matsubara2020deep, Jin2020sympnets, David2021symplectic, Eidnes2022port, Noren2023learning}, where we use a loss function evaluated on the integration scheme:
\begin{equation}\label{eq:lossbarr}
\mathcal{L} = \left\lVert  \frac{x^{n+1}-x^{n}}{\Delta t} - \Psi_{\Delta t}(\hat{g}_\theta,x^{n},x^{n+1},t^n) \right\rVert_2,
\end{equation}
given for one data point $x^n$ and barring regularization.

By contrast, works like \cite{Chen2018neural, Chen2020Symplectic, DiPietro2020sparse, Desai2021variational} use a loss function evaluating the difference between a data point and what is obtained doing integration:
\begin{equation}\label{eq:integrationloss}
\mathcal{L} = \lVert  x^{n+1} - \hat{x}^{n+1}\rVert_2 =\lVert  x^{n+1} - (\hat{x}^n + \Delta t \Psi_{\Delta t}(\hat{g}_\theta,\hat{x}^n,\cdot,t^n))\rVert_2.
\end{equation}
Note that the approaches are equivalent if the same explicit integrator is used and only one integration step is done between the data points so that $\hat{x}^n=x^n$ in \eqref{eq:integrationloss}.

When wishing to approximate the ODE \eqref{eq:ode} using two available successive data points $x^n=x(t^n)$ and $x^{n+1}=x(t^n+\Delta t)$, we say that a discretization method \eqref{eq:integrationscheme} is explicit if $\Psi_{\Delta t}(g,x^n,x^{n+1},t^n)$ depends explicitly only on the two points $x^n$ and $x^{n+1}$ in addition to $t^n$ and $\Delta t$, i.e.\ without relying on implicit intermediate steps. Note that this is different from the definition of an explicit integrator, which cannot depend on $x^{n+1}$. This class of explicit discretization methods includes the methods that lead to explicit integrators, but also implicit integrators that do not require numerical integration to obtain intermediate steps, like the implicit midpoint method \eqref{eq:midpoint}. 
Within the larger class of Runge--Kutta methods, explicit discretization methods correspond to the class called mono-implicit Runge--Kutta methods \cite{vanBokhoven1980efficient, Cash1982mono}. This class does not include higher-order symplectic methods like the Gauss--Legendre methods of orders four and six, which require intermediate steps to be found implicitly.

For our numerical experiments, we have used the fourth-order symmetric integrator first proposed in \cite{Eidnes2022port}, where it is formally given for time-independent $g$. Also including time-dependency, the method is given by
\begin{equation}\label{eq:srk4}
\begin{split}
\Psi_{\Delta t}&(g, x^n,x^{n+1},t^n) = \\
& \,
\frac{1}{2} \, g\left( \frac{x^{n}+x^{n+1}}{2}-\frac{\sqrt{3}}{6} \Delta t \, g( c_1 x^n + c_2 x^{n+1}, t^n + c_2\Delta t), \, t^n + c_1\Delta t\right)\\
& \, + \frac{1}{2} \, g\left( \frac{x^{n}+x^{n+1}}{2}+\frac{\sqrt{3}}{6} \Delta t \, g( c_2 x^n + c_1 x^{n+1}, t^n + c_1\Delta t), \, t^n + c_2\Delta t\right),
\end{split}
\end{equation}
for $c_1 = \frac{1}{2} - \frac{\sqrt{3}}{6}$ and $c_2 = \frac{1}{2} + \frac{\sqrt{3}}{6}$. 
An equivalent definition can be represented by the Butcher tableau
\begin{equation*}
\renewcommand\arraystretch{1.2}
\begin{array}
{c|cccc}
\frac{1}{2}-\frac{\sqrt{3}}{6} & \frac{1}{4} & 0 & -\frac{\sqrt{3}}{6} &\frac{1}{4}  \\
\frac{1}{2}-\frac{\sqrt{3}}{6} & \frac{1}{4}-\frac{\sqrt{3}}{12}& 0  & 0 &\frac{1}{4}-\frac{\sqrt{3}}{12}  \\
\frac{1}{2}+\frac{\sqrt{3}}{6} & \frac{1}{4}+\frac{\sqrt{3}}{12}& 0  & 0 &\frac{1}{4}+\frac{\sqrt{3}}{12}  \\
\frac{1}{2}+\frac{\sqrt{3}}{6} & \frac{1}{4} & \frac{\sqrt{3}}{6} & 0 &\frac{1}{4} \\
\hline
&\frac{1}{2} & 0 & 0 & \frac{1}{2} 
\end{array}
\end{equation*}
We see that it satisfies the conditions for fourth order, as given in Table III.1.1 of \cite{Hairer06} (since all symmetric methods have even order, it is sufficient to check the third order conditions). We see also that it does not satisfy the conditions for symplecticity, as given in Theorem VI.4.3 of \cite{Hairer06}.

Several papers have suggested using symplectic integrators for learning Hamiltonian systems \cite{Chen2020Symplectic, DiPietro2020sparse, Desai2021variational}, but a solid theoretical argument for this has to our understanding not yet been provided. In numerical integration of a Hamiltonian system, symplectic integrators are often viewed as favorable for long-term integration, since they guarantee exact preservation of an approximated Hamiltonian, and preservation of the exact Hamiltonian within a bound \cite{Hairer06}. This provides qualitative properties, e.g.\ a guarantee for stability, that do not translate directly to the inverse problem of learning a system from known data.

The integrator used in \cite{DiPietro2020sparse} is Yoshida's fourth-order method, which is also suggested for HNN in \cite{Desai2021variational}, although neither paper refers to it by that name or cite the proper reference \cite{Yoshida1990construction}. This method is a partitioned Runge--Kutta method that is explicit for separable systems, but neither explicit nor mono-implicit for non-separable systems. Thus it would not be a good choice if one cannot assume the system to be separable, but \cite{DiPietro2020sparse, Desai2021variational} do assume separability and argue for using this integrator because it is a fourth-order symplectic method. An advantage of using \eqref{eq:integrationloss} with an explicit integrator is that one can do several steps between each training data point without having to rely on a root-finding algorithm like Newton's method, and \cite{DiPietro2020sparse} do indeed take advantage of this to do several shorter integration steps for each training step. This property means that given noise-free data, one can achieve arbitrarily high accuracy on the integration from $x^n$ to $\hat{x}^{n+1}$, limited only by computational resources and time. However, it is generally less computationally efficient to use many steps than to use a higher-order integrator, and in the presence of noise, using more steps does not alleviate the issue of getting the noise from only one data point.

\section{Numerical comparison of integration schemes}
We set up PHSI models with different integrators in the loss function \eqref{loss} and everything else the same, to test how well different integrators with different properties affect the results. The integrators are
\begin{itemize}
    \item the forward Euler method (Euler)
    \item the implicit midpoint method \eqref{eq:midpoint} (Midpoint)
    \item the classic fourth-order Runge--Kutta method (RK4)
    \item the symmetric fourth-order Runge--Kutta method \eqref{eq:srk4} (SRK4)
    \item Cash and Singhal's symmetric sixth-order Runge--Kutta method \cite{Cash1982high} (SRK6)
    \item Yoshida's fourth order partitioned Runge--Kutta method (PRK4)
\end{itemize}
and their properties are summarized in Table \ref{tab:integrators}.

\begin{table}[ht!]
\centering
\caption{Properties of integrators. PRK4 is explicit, mono-implicit and symplectic for separable systems but not for non-separable systems.}
\label{tab:integrators}
\begin{tabular}{lcccccc}
\toprule
Integrator & order & $g$ eval's & explicit & mono-implicit & symmetric & symplectic \\ 
\midrule
Euler & $1$ & $1$ & yes & yes & no & no \\
Midpoint & $2$ & $1$ & no & yes & yes & yes \\
RK4 & $4$ & $4$ & yes & yes & no & no \\
SRK4 & $4$ & $4$ & no & yes & yes & no \\
SRK6 & $6$ & $5$ & no & yes & yes & no \\
PRK4 & $4$ & $7$ & yes/no & yes/no & yes & yes/no \\
\bottomrule
\end{tabular}
\end{table}

\subsection{Hamiltonian system: H\'enon--Heiles}

We train the models on trajectories from $t=0$ to $t=10$ with
\begin{itemize}
    \item sampling time $1$ and $250$ samples, i.e.\ $25$ trajectories, without noise;
    \item sampling time $1$ and $250$ samples with moderate noise (Gaussian noise with a standard deviation $\sigma = 0.03$ added to the measurements of the states);
    \item sampling time $1$ and $250$ samples with much noise (standard deviation $\sigma = 0.05$);
    \item sampling time $1/2$ and $1000$ samples, i.e.\ $50$ trajectories, without noise;
    \item sampling time $1/2$ and $1000$ samples with moderate noise;
    \item sampling time $1/2$ and $1000$ samples with much noise.
\end{itemize}

We generally see a better performance from the symmetric mono-implicit Runge--Kutta methods (Midpoint, SRK4, SRK6) than those that only rely on one data point in the evaluation of $\Psi_{\Delta t}$. This seems to define a more important property than symplecticity; the symplectic PRK4 method performs consistently worse than the non-symplectic SRK4 method. The implicit midpoint method is the only method that is both symplectic and relies equally on $x^n$ and $x^{n+1}$, and it does perform well despite being only a second-order method.

\begin{figure}[ht!]
        \centering
        \includegraphics[scale=1.]{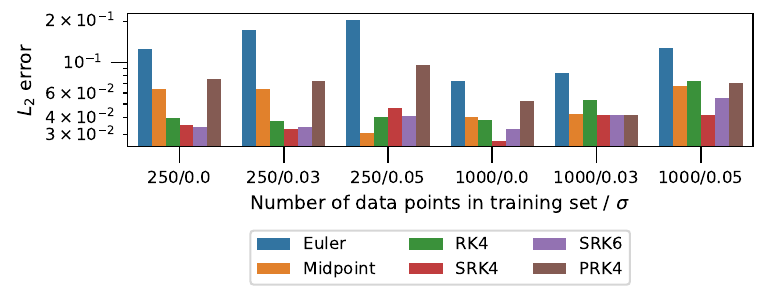}
    \caption{The mean $L_2$ error of PHSI models trained with the different integrators on the H\'enon--Heiles problem. The error is on the predicted positions and momenta from $t=0$ to $t=10$ on 10 different random initial conditions.}
\label{fig:hh_integrators_barplot}
\end{figure}

Note that while we train on the integration scheme in a way that is equivalent to performing one integration step, DiPietro et al.\ \cite{DiPietro2020sparse} perform several steps of the PRK4 integrator between data points to increase accuracy. We assume that this is a major reason for the superior performance over using the RK4 method reported in that paper, and also note that a major improvement in performance by taking many smaller integration steps will almost only happen if one trains on noise-free data.

\begin{figure}[ht!]
        \centering
        \includegraphics[scale=1.]{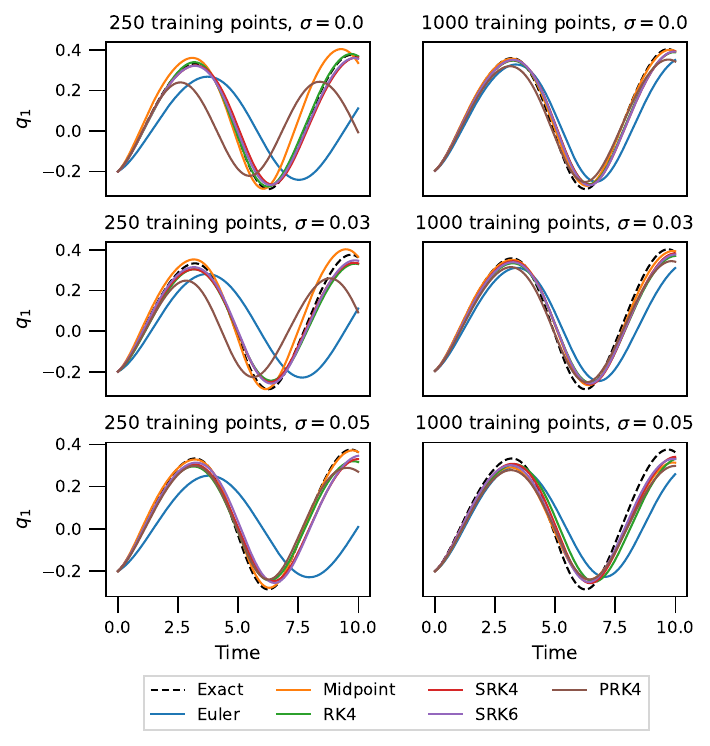}
        \caption{Prediction of $q_1$ by PHSI models trained with different integrators on the H\'enon--Heiles problem. The initial condition is $q=(-0.2, 0.2)$, $p=(0.1,-0.2)$.}
\label{fig:hh_integrators_trajectories}
\end{figure}

\subsection{Pseudo-Hamiltonian system: Tanks and pipes}

We also test the performance of the different integrators in the hybrid model used to learn the tank system presented in Section \ref{subsec:hybrid}. We do not test Yoshida's 4th order method on this problem, since the tank system is not a separable partitioned system, and so Yoshida's method is in this case an implicit method that would require a root-finding algorithm like Newton's method to be used in every iteration of the training process. We have not implemented this in our models, as it would be prohibitively expensive.

We train the models on trajectories from $t=0$ to $t=1$ with
\begin{itemize}
    \item sampling time $1/50$ and $7500$ samples, i.e.\ $300$ trajectories, without noise;
    \item sampling time $1/50$ and $7500$ samples with moderate noise (Gaussian noise with a standard deviation $\sigma = 0.03$ added to the measurements of the states);
    \item sampling time $1/50$ and $7500$ samples with much noise (standard deviation $\sigma = 0.05$);
    \item sampling time $1/100$ and $30000$ samples, i.e.\ $150$ trajectories, without noise;
    \item sampling time $1/100$ and $30000$ samples with moderate noise;
    \item sampling time $1/100$ and $30000$ samples with much noise.
\end{itemize}

\begin{figure}[ht!]
        \centering
        \includegraphics[scale=1.]{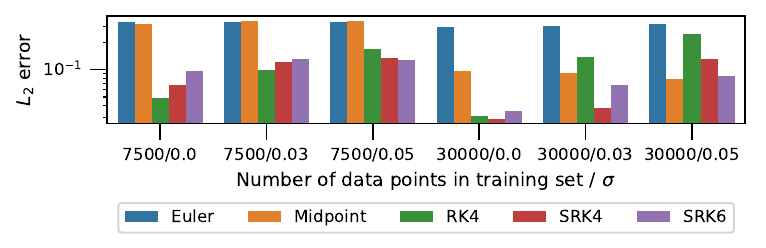}
    \caption{The mean $L_2$ error of hybrid PHSI models trained with the different integrators on the tank system. The error is of the predicted volume and flow in all tanks and pipes from $t=0$ to $t=1$ on 10 different random initial conditions.}
\label{fig:tank_integrators_barplot}
\end{figure}

\begin{figure}[ht!]
        \centering
        \includegraphics[scale=1.]{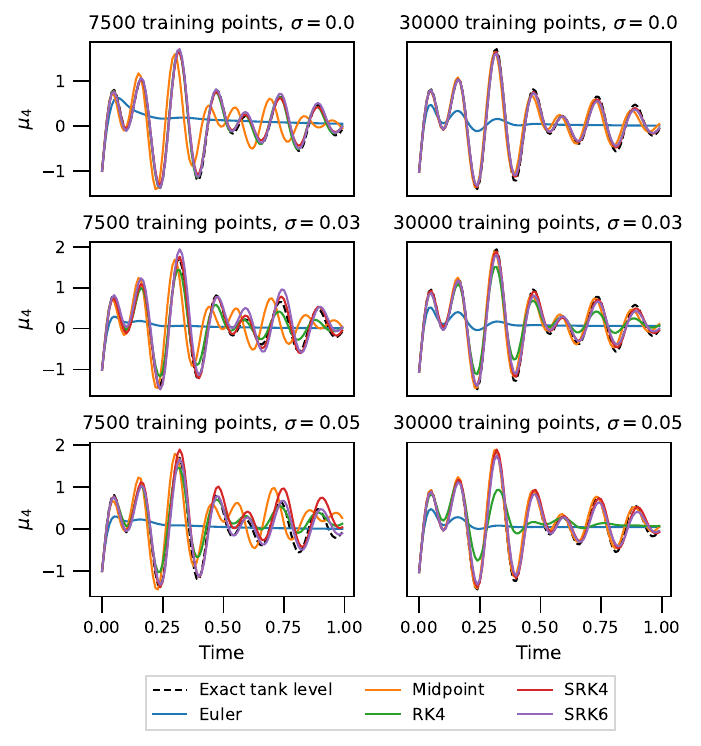}
        \caption{Volume of the fourth tank as predicted by hybrid PHSI models trained with different integrators on the tank system. The initial condition is $\phi^0 = (-1, -1, 0, \frac{1}{2}, -1)$, $\mu^0= (1, 1, -\frac{1}{2}, -1)$.}
\label{fig:tank_integrators_trajectories}
\end{figure}

Figure \ref{fig:tank_integrators_barplot} shows the $L_2$ error of all the state variables from the predictions obtained applying the methods on a test set of 10 different initial conditions sampled from independent uniform distributions $\mathcal{U}(-1, 1)$. The symmetric higher-order methods are more consistent in their performance than the lower-order methods and the explicit fourth-order Runge--Kutta method. As for the H\'enon--Heiles system, we see no consistent improvement in going from fourth to sixth order, which supported the choice of using the SRK4 method throughout this paper. The symmetric methods particularly predict the tank volumes very well, more so than the pipe flows. An example is shown for the volume in the fourth tank, i.e.\ the one with a leak, in Figure \ref{fig:tank_integrators_trajectories}.

Table \ref{tab:integrators_R} shows how well the friction coefficients on the pipes are learned by the different integrators. We see here, even clearer than on the $L_2$ error, that the symmetric methods perform better than the explicit fourth-order Runge--Kutta method, particularly on noisy data.

\begin{table}[ht!]
\centering
\caption{Mean and standard deviation of the predicted friction coefficients of the tank system, relative to the ground truth $R_p = (0.03, 0.03, 0.09, 0.03, 0.03)$ (i.e.\ so that $1$ corresponds to the correct coefficient).}
\label{tab:integrators_R}
\begin{tabular}{lccc}
\toprule
& no noise & $\sigma=0.03$ & $\sigma=0.05$ \\
\midrule
\multicolumn{4}{c}{7500 training points} \\
\midrule
Euler & $22.75\pm7.20$ & $23.59\pm7.04$ & $25.27\pm7.31$ \\
Midpoint & $1.27\pm0.04$ & $1.16\pm0.14$ & $1.01\pm0.11$ \\
RK4 & $1.04\pm0.04$ & $1.70\pm0.24$ & $1.95\pm0.27$ \\
SRK4 & $1.16\pm0.07$ & $0.92\pm0.13$ & $0.88\pm0.43$ \\
SRK6 & $0.95\pm0.07$ & $0.85\pm0.19$ & $0.94\pm0.25$ \\
\midrule
\multicolumn{4}{c}{30000 training points} \\
\midrule
Euler & $11.33\pm3.31$ & $12.31\pm3.48$ & $13.83\pm3.91$ \\
Midpoint & $1.23\pm0.11$ & $1.15\pm0.16$ & $1.16\pm0.25$ \\
RK4 & $1.20\pm0.06$ & $1.97\pm0.29$ & $4.18\pm0.70$ \\
SRK4 & $1.16\pm0.04$ & $1.11\pm0.13$  & $1.00\pm0.23$ \\
SRK6 & $1.24\pm0.06$ & $1.21\pm0.10$  & $1.35\pm0.32$ \\
\bottomrule
\end{tabular}
\end{table}

\section{A study of promoting sparsity through regularization and pruning}
In sections \ref{sec:pruning} and \ref{sec:regularization}, we argue that regularization and pruning will help the PHSI model promote sparsity in search of the true governing equations. It is also argued that since we know that the true governing equations are generally sparse in the function space, promoting sparsity during training will help the PHSI model better approximate the true system. To evaluate whether this claim has merit or not, we train PHSI models where the amounts of regularization and pruning vary and the rest of the hyperparameters are held constant. By the "amount" of pruning, we mean how often the pruning algorithm is employed during training, i.e.\ the value $P$ in Algorithm \ref{training_alg}. The lower value $P$ has, the more frequent the pruning. By the "amount" of regularization, we specifically mean the amount of $l_1$-regularization on the Hamiltonian $\hat{H}$, i.e. $\lambda_H$ in (\ref{loss}). We evaluate the predictive performance of each trained model by comparing their "predictability score", i.e. their average error over simulated trajectories with random initialization compared to the true trajectories, where the trajectories are not from the training set. The tests are done for all of the four systems considered in the Section \ref{sec:experiments}, and they are performed on noiseless data. 

\subsection{Experiment 1: Hénon--Heiles system}
\begin{figure}[ht!]
    \centering
    \includegraphics[scale=0.6]{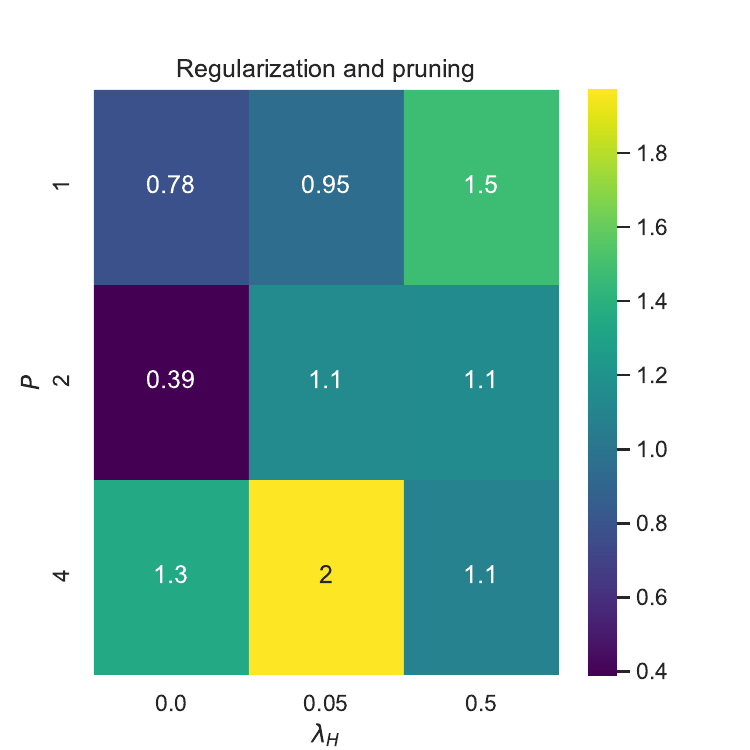}
    \caption{Average $L_2$-error over 30 simulated trajectories of the Hénon--Heiles system by PHSI models trained with different combinations of regularization parameter $\lambda_H$ and pruning parameter $P$. The models are trained for $3$ epochs, which means that when $P=4$, there is no pruning.}
    \label{fig:pruning_hh_heatmap}
\end{figure}
\begin{figure}[ht!]
    \centering
    \includegraphics[width=\textwidth]{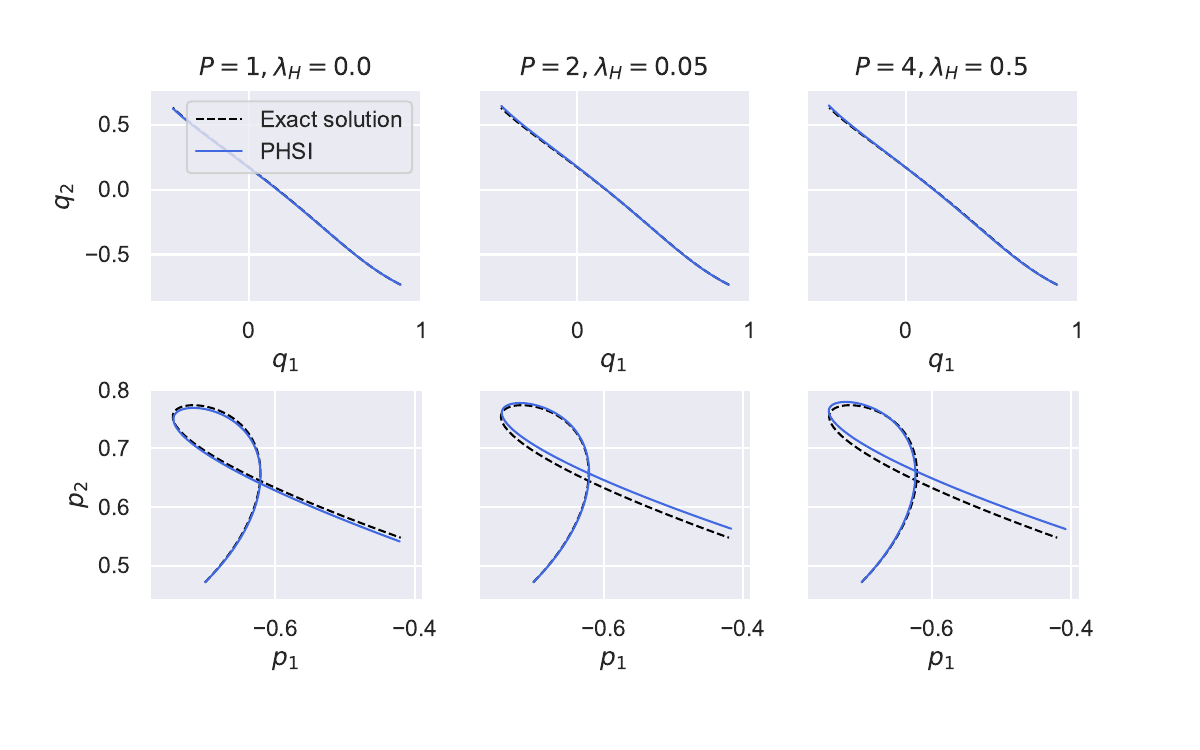}
    \caption{Simulated trajectories obtained for different PHSI models and the exact solution, for the nonlinear Hénon--Heiles system for three different combinations of regularization and pruning. }
    \label{fig:pruning_hh_trajectory}
\end{figure}
The Hénon--Heiles system is a purely Hamiltonian system, and this is assumed when training the PHSI model. The models are trained on noiseless, simulated data with $\lambda_H \in \{0, 0.05, 0.5\}$ and $P\in \{1, 2, 4\}$, for $3$ epochs. The predictive performances of models with different combinations of regularization and pruning are shown in Figure \ref{fig:pruning_hh_heatmap}. The figure shows that for this particular system, the PHSI model trains best without $l_1$-regularization, and hence no regularization was used in the experiments in Section \ref{sec:experiments}. The reason for this may lie in that the PHSI model learns the relatively simple Hénon--Heiles system in very few epochs of training. The pruning algorithm does not appear to affect the average $L_2$-error in this experiment. However, it excludes the correct terms, resulting in easier interpretability of the trained model. Figure \ref{fig:pruning_hh_trajectory} shows that although the scores in Figure \ref{fig:pruning_hh_heatmap} differ by several orders of magnitude, all the models achieve a relatively good predictive ability. 

\subsection{Experiment 2: Nonlinear Schrödinger system}
\begin{figure}[ht!]
    \centering
    \includegraphics[scale=0.6]{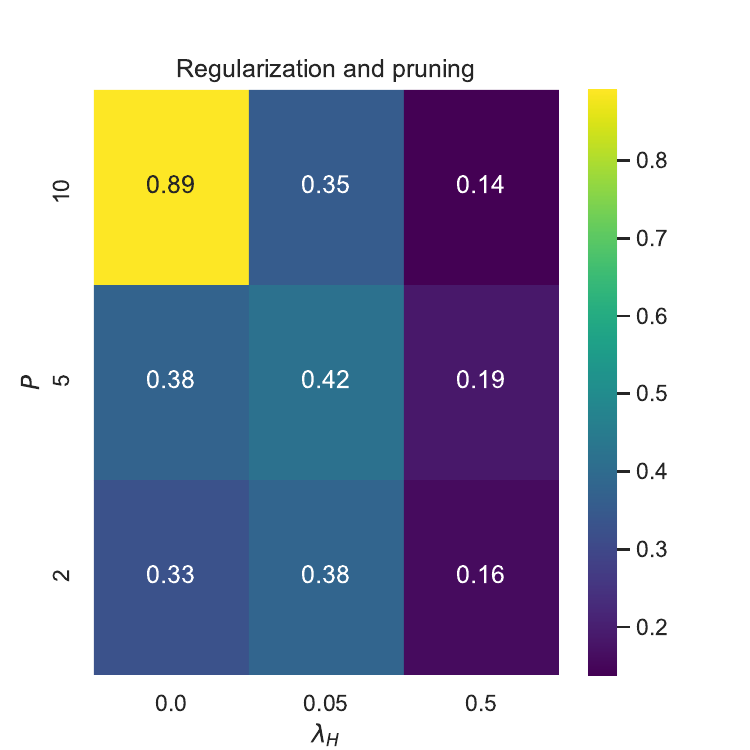}
    \caption{Average $L_2$-error over 30 simulated trajectories of the nonlinear Schrödinger system by PHSI models trained with different combinations of regularization parameter $\lambda_H$ and pruning parameter $P$. The models are trained for $10$ epochs, meaning when $P=10$, the pruning only occurs once.}
    \label{fig:pruning_nls_heatmap}
\end{figure}

\begin{figure}[ht!]
    \centering
    \includegraphics[width=\textwidth]{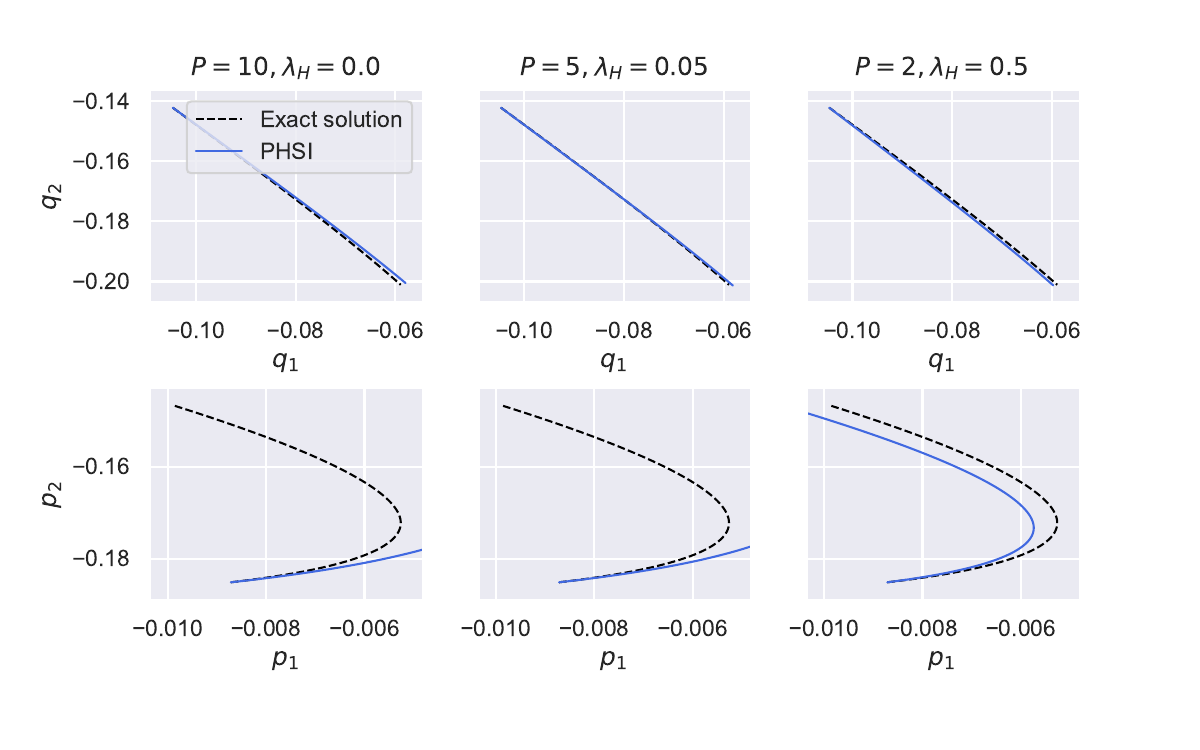}
    \caption{Simulated PHSI trajectories along with a simulation of the exact nonlinear Schödinger system, for three different combinations of regularization and pruning. }
    \label{fig:pruning_nls_trajectory}
\end{figure}

The non-linear Schrödinger system is also purely Hamiltonian, and this is assumed when training the PHSI model, in that we do not train damping coefficients or an external force. The models are trained with $\lambda_H \in \{0, 0.05, 0.5\}$ and $P\in \{1, 5, 10\}$, for $10$ epochs. In Figure \ref{fig:pruning_nls_heatmap}, we observe that the regularization improves the model performance drastically. The model with $P=10$ and $\lambda_H = 0$ has an average $L_2$-error approximately 5 times that of the model with $P=2$ and $\lambda_H = 0.5$, suggesting that the regularization and pruning work well together. This effect being so evident here and not for the Hénon--Heiles system may be due to the non-linear Schrödinger system being non-separable and of a higher polynomial order, making the true solution more sparse in the function space. Figure \ref{fig:pruning_nls_trajectory} illustrates the vastly superior performance of the models with more regularization and pruning over those with less.


\subsection{Experiment 3: Damped mass-spring system}
\begin{figure}[ht!]
    \centering
    \includegraphics[scale=0.6]{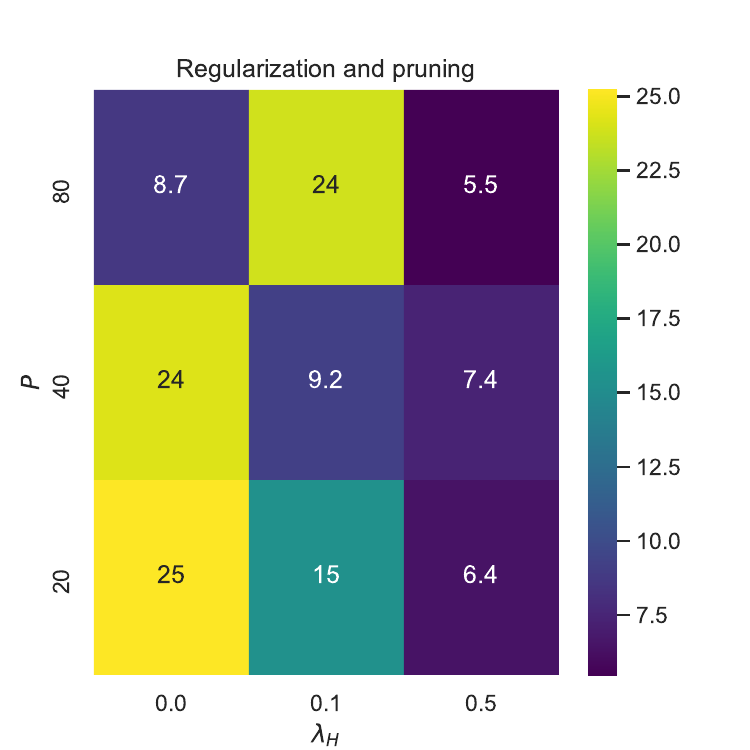}
    \caption{Average $L_2$-error over 30 simulated trajectories of the damped mass-spring system by PHSI models trained with different combinations of regularization-parameter $\lambda_H$ and pruning-parameter $P$. The models are trained for $80$ epochs, so that when $P=80$, pruning is only done after the last epoch.}
    \label{fig:pruning_msd_heatmap}
\end{figure}
\begin{figure}[ht!]
    \centering
    \includegraphics[width=\textwidth]{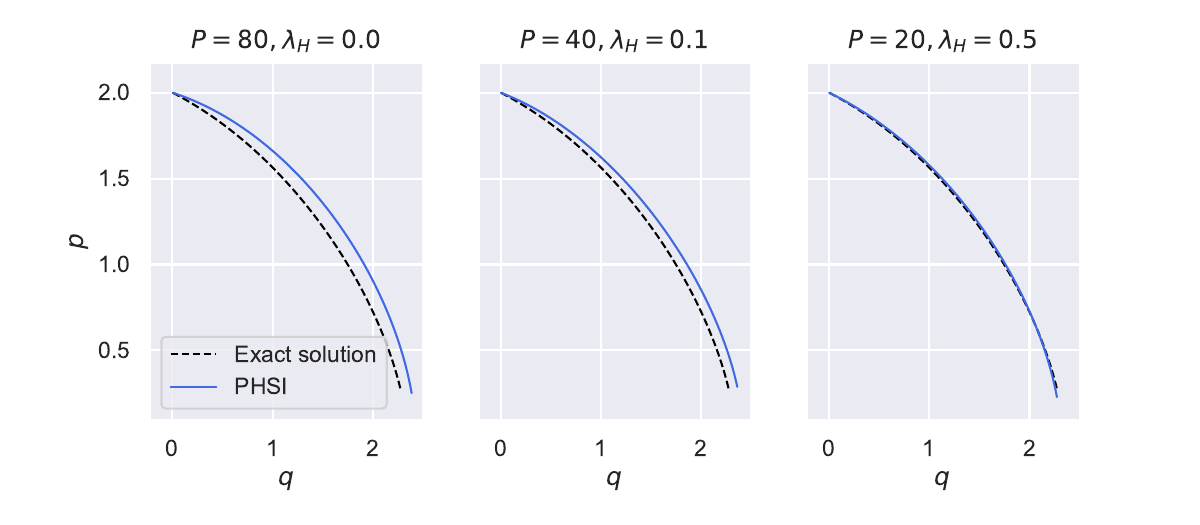}
    \caption{Simulated PHSI trajectories along with a simulation of the exact damped mass-spring system, for three different combinations of regularization and pruning. }
    \label{fig:pruning_msd_trajectory}
\end{figure}
The damped mass-spring system described in Section \ref{subsec:msd} is a pseudo-Hamiltonian system. We have regularization on the Hamiltonian and the external forces, i.e.\ $\lambda_H$ and $\lambda_F$ in \eqref{loss}, but not on the damping coefficients. The models are trained for $80$ epochs, with $\lambda_H \in \{0, 0.1, 0.5\}$ and $P\in \{20, 40, 80\}$. $\lambda_F$ is set to $0.01$. Figure \ref{fig:pruning_msd_heatmap} indicates that promoting sparsity helps the predictive performance of the PHSI model, especially through $l_1$-regularization. The superior predictive ability of the PHSI models that promote sparsity is also seen in the plotted simulated trajectories in Figure \ref{fig:pruning_msd_trajectory}. 

\subsection{Experiment 4: Tank system}
\begin{figure}[ht!]
    \centering
    \includegraphics[scale=0.6]{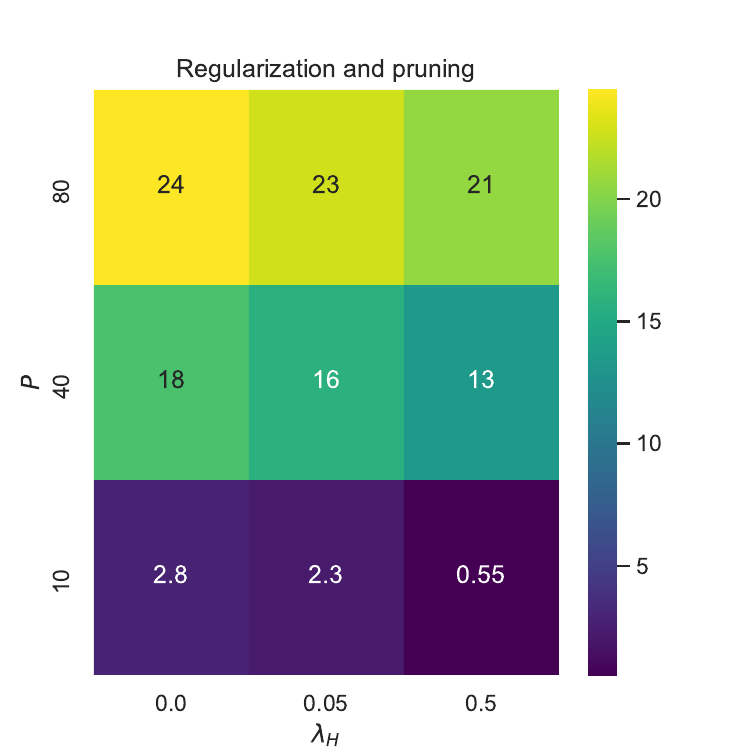}
    \caption{Average loss over simulated trajectories for PHSI models trained with different combinations of regularization-parameter $\lambda_H$ and pruning-parameter $P$. The models are trained for $80$ epochs, which means that when $P=80$, pruning is only done after the last epoch.}
    \label{fig:pruning_heatmap}
\end{figure}

For the tank system considered in Section \ref{subsec:hybrid}, we choose the search-spaces $\lambda_H \in \{0, 0.05, 0.5\}$ and $P \in \{10, 40, 80\}$, and train the models for 80 epochs. As shown in Figure \ref{fig:pruning_heatmap}, the regularization and pruning have a great impact on the performance of the model, and appear essential for achieving an accurate PHSI model. The model with $\lambda_H = 0$ and $P=80$ has a predictive score about 60 times that of the model with $\lambda_H = 0.5$ and $P=10$.  Figure \ref{fig:pruning_heatmap} confirms that this assumption is true for the tank system. 
\begin{figure}[ht!]
    \centering
    \includegraphics[width=\textwidth]{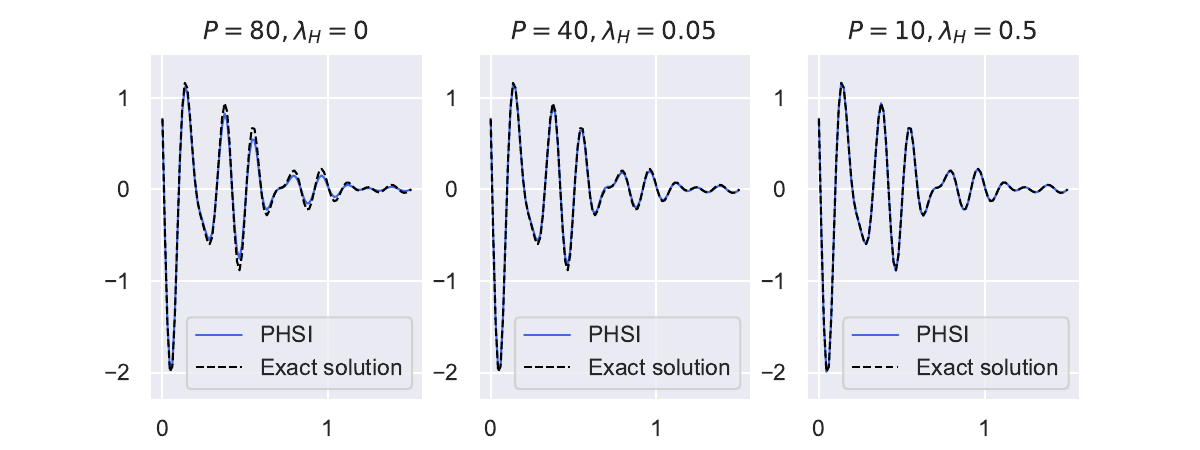}
    \caption{Simulated PHSI trajectories along with the exact solution of one of the leaking tanks in the connected tank system for three different combinations of regularization and pruning.}
    \label{fig:pruning_tank_trajectory}
\end{figure}

In conclusion, promoting sparsity in the function space generally improves the accuracy of the PHSI model. Since promoting sparsity in the PHSI model through regularization and pruning improved results for three of the four models (it neither improved nor worsened the performance when learning the Hénon--Heiles system), the sparsity assumption made in sections \ref{sec:pruning} and \ref{sec:regularization} seems to be a good one. This again implies that the PHSI model is in general well fit for finding true governing equations. The experiments indicate that the regularization and pruning has more of an impact when learning systems of high dimensionality and polynomial order. If the PHSI model is to be applied to learning more complex and higher dimensional systems than what has been done in this thesis, promoting sparsity will presumably be vital for achieving an accurate model. 

\section{A study of regularization: learning a mass-spring system} \label{appendix_regularization}
When learning dynamical systems of complex structures, regularization is important not only to promote sparsity but to ensure the desired structure of the ODE. We exemplify this by studying a mass-spring system, without damping or external forces, and learning it through a pseudo-Hamiltonian structure \eqref{port_ham_eq}. A simple-mass spring system can be described through the following ODE:
\begin{equation}
    \begin{bmatrix} \dot{q} \\ \dot{p} \end{bmatrix} 
    = \begin{bmatrix} \frac{p}{m} \\ -kq \end{bmatrix}.
\end{equation}
We set weight $m=1$ and spring stiffness $k=1$. The system then has the related Hamiltonian function
\begin{equation}
    H(q, p) = \frac{1}{2}q^2 + \frac{1}{2} p^2,
\label{eq:example_h}
\end{equation}
which we can use to write the ODE as a Hamiltonian system with the canonical formulation
\begin{equation}
    \begin{bmatrix} \dot{q} \\ \dot{p} \end{bmatrix}
    = \begin{bmatrix} 0 & 1 \\ -1 & 0 \end{bmatrix}
    \begin{bmatrix} \frac{\partial H}{\partial q} \\ \frac{\partial H}{\partial p} \end{bmatrix}.
\end{equation}
The model will be trained assuming a pseudo-Hamiltonian formulation (\ref{port_ham_eq}) without damping, where the Hamiltonian and the external forces are trainable system identification models depending on the state only and the structure matrix is known:
\begin{equation}
    \hat{g}_\theta(x) = \begin{bmatrix} 0 & 1 \\ -1 & 0 \end{bmatrix} \nabla \hat{H}_\theta(x) + \hat{F}_\theta(x).
\end{equation}
As previously stated, the pseudo-Hamiltonian formulation is non-unique since a general ODE can be equivalently represented by different combinations of $H$ and $F$. Since system identification models attempt not only to accurately predict future states but identify the governing equations, we are interested in separating $\hat{H}_\theta$ and $\hat{F}_\theta$ in a natural way. In this example, since there are no external forces present, the desired solution is $\hat{H}_\theta$ as given by (\ref{eq:example_h}) and $\hat{F}_\theta = 0$.
\begin{figure}[ht!]
    \centering
    \includegraphics[scale=0.6]{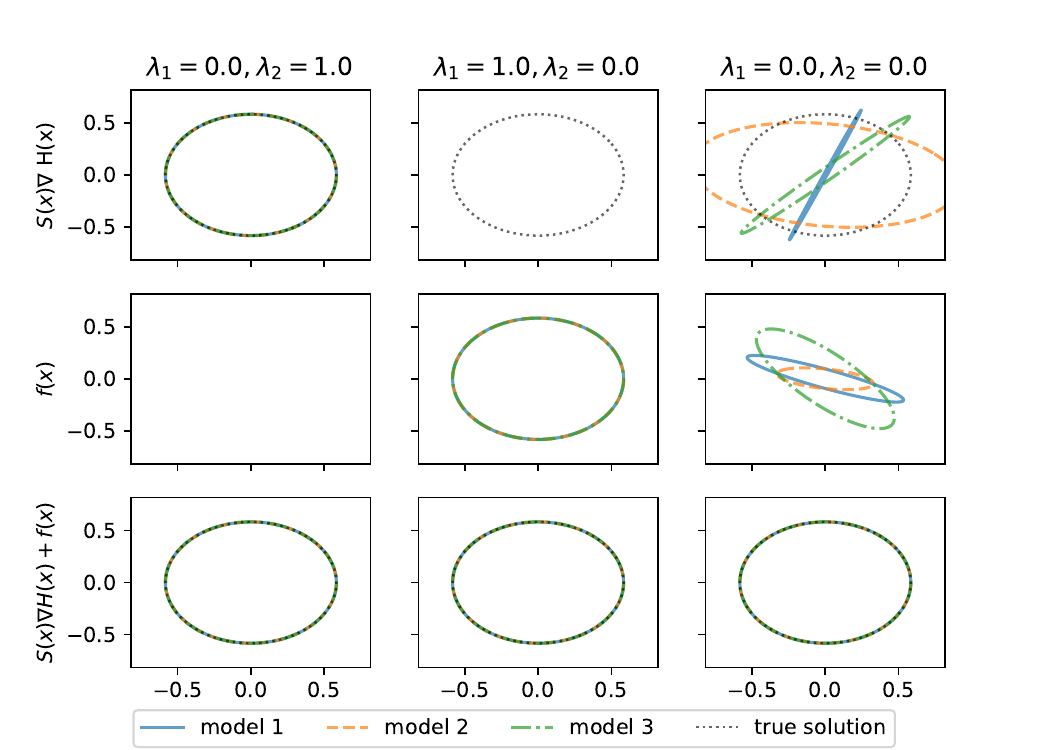}
    \caption{Phase plots of trajectories $x=(q,p)$ obtained from systems where $\hat{\dot{x}}$ is given by different combinations of $S(x)\nabla H$ and $f$. Each column is a different combination of regularization parameters, where $\lambda_1$ and $\lambda_2$ are the magnitudes of $l_1$ regularization on $\hat{H}_\theta$ and $\hat{F}_\theta$, respectively.}
    \label{fig:msdexample_plot}
\end{figure}
We train system identification models with pseudo-Hamiltonian structures with different combinations of regularization on $\hat{H}_\theta$ and $\hat{F}_\theta$. Figure \ref{fig:msdexample_plot} shows trajectory plots of three randomly initiated models compared to the ground truth. Every model trained predicts $\hat{\dot{x}}$ accurately, but only the models that regularize $\hat{F}_\theta$ are able to separate $\hat{H}_\theta$ and $\hat{F}_\theta$ in the desired way. The models that regularize $\hat{H}_\theta$ learn the entire ODE through $\hat{F}_\theta$, and the models without regularization learn difficult-to-interpret combinations between $\hat{H}_\theta$ and $\hat{F}_\theta$ that are non-unique due to the random initialization of the model parameters. These results show how regularization affects the structure of a trained PHSI model. How we set the regularization parameters of a PHSI model reflects the prior assumptions we make about the system to be learned.

\end{document}